\documentclass[aps,twocolumn,prc,superscriptaddress,showpacs,nofootinbib,floatfix,amssymb,amsfonts,amsmath]{revtex4-1}

\usepackage{graphicx}
\usepackage{epsf}
\usepackage{amsmath,amssymb}
\usepackage{bm}
\usepackage{color}
\usepackage{ulem}

\newcommand{\cent}{c}

\begin{document}

\title{
Challenge for describing the cluster states starting with realistic interaction
}

\author{Naoyuki Itagaki}
\author{Tokuro Fukui}
\affiliation{
Yukawa Institute for Theoretical Physics, Kyoto University,
Kitashirakawa Oiwake-Cho, Kyoto 606-8502, Japan
}

\author{Akihiro Tohsaki}
\affiliation{
Research Center for Nuclear Physics (RCNP), Osaka University,
10-1 Mihogaoka, Ibaraki, Osaka 567-0047, Japan
}

\date{\today}

\begin{abstract}
We aim to describe the cluster states of nuclear systems starting with a realistic interaction,
which is a challenge of the modern nuclear theories.
Here, the short-range correlation of realistic interaction  is treated by employing the damping factor,
and the resultant interaction can be applied to the cluster structure of light nuclei.
We start with a realistic interaction (G3RS)
and transform it in this way, and the $\alpha$-$\alpha$ energy curve is compared with
the results of phenomenological interactions.
The attractive effect between two  $\alpha$'s is found to be not enough even with 
a damping factor for the short-range repulsion, and the necessity of a finite-range three-body term
is discussed.
With this three-body term, 
the resonance energy of the ground state and  the scattering phase shift of two $\alpha$'s can be reproduced.
Also, the binding energy of $^{16}$O from the four $\alpha$ threshold is reasonably reproduced.
The linear-chain structure of three and four $\alpha$ clusters
in $^{12}$C and $^{16}$O
are calculated with this interaction 
and compared with the results of the conventional approaches
including the density functional theories. 
\end{abstract}

\maketitle

\section{Introduction}
Describing the cluster states starting with the realistic interactions is a challenge of the modern nuclear theories.
The $^4$He nucleus is a strongly bound many-nucleon system
in the light mass region, thus the $\alpha$ clusters
can be basic building blocks of the nuclear structure.
The $\alpha$ cluster models~\cite{Brink,PTPS.68.29} have been developed and applied 
in numerous works
for the description of 
cluster structures such as $3\alpha$ clustering in
the so-called Hoyle state of $^{12}$C~\cite{Hoyle,Uegaki12C,PhysRevLett.87.192501}.
In most of the conventional
cluster models, each $\alpha$ cluster is assumed as 
a simple $(0s)^4$ configuration.
However, in the real systems,
nucleons are correlated owing to the repulsive core in the short-range part of the central interaction,
and this effect is not explicitly treated in the conventional cluster models. 
These days, such $NN$ correlation is widely discussed 
based on modern {\it ab initio} theories not only in very light nuclei
but also in medium-heavy nuclei~\cite{BINDER2014119}.
Most of the  {\it ab initio} theories have been developed based on the shell-model
point of view, and 
describing cluster states
is a challenge of
the modern $ab\ initio$ ones~\cite{PhysRevC.79.014308,DREYFUSS2013511,Yoshida_2014},
since a quite large model space is required.

Concerning the description of the cluster states starting with 
realistic interactions, one of the ways is to utilize
Fermionic molecular dynamics (FMD) combined with the 
unitary correlation operator method (UCOM)~\cite{NEFF2004357,PhysRevLett.105.022501}.
In UCOM, 
the effects of the short-range correlation are
included with the unitary transformation of the Hamiltonian,
which 
in principle induces many-body operators up to the $A$-body level, with the mass number $A$
since we need to expand this unitary operator for the calculation of the expectation values of the norm and the Hamiltonian.
We aim
to treat the short-range correlation 
caused by the repulsive core of the central interaction in a simple way
without performing the unitary transformation and expanding the exponents.

The assumption of the
$(0s)^4$ configuration for the $\alpha$ clusters in the conventional $\alpha$ cluster models
also
prevents us to include the contribution of the non-central interaction.
We have previously introduced 
antisymmetrized quasi cluster model (AQCM)~\cite{PhysRevC.83.014302,ptep093D01,PhysRevC.87.054334,PhysRevC.94.064324,PhysRevC.97.014307,PhysRevC.98.044306,PhysRevC.71.064307,PhysRevC.75.054309} and 
tensor version of AQCM
(AQCM-T)~\cite{PTEP-Itagaki-ptz046,PhysRevC.98.054306,PhysRevC.73.034310,PhysRevC.97.014304} to include the contribution of the non-central interactions
in the $\alpha$-cluster model by breaking the $(0s)^4$ configuration properly. 
It has been discussed that the tensor suppression effect 
at small $\alpha$-$\alpha$ distances
works repulsively and
plays an important role in the appearance
of the $\alpha$ cluster structures~\cite{PTEP-Itagaki-ptz046,PhysRevC.98.054306,PhysRevC.97.014304}.
In this paper, however, we do not touch into the non-central interactions, 
since we are interested only in the short-range correlation originating from 
the central interaction. Therefore, here 
the model space is the conventional Brink model with the $(0s)^4$ configuration for each $\alpha$ cluster.
In return, we start with a realistic interaction, G3RS
(Gaussian three-range soft-core potential)~\cite{PTP.39.91}
for the central part.

Until now, in conventional cluster model studies, 
phenomenological interactions, such as
Volkov interaction~\cite{VOLKOV196533}, have been
widely used, 
which allows us to reproduce the $\alpha$-$\alpha$ scattering phase shift
by 
properly choosing
the Majorana exchange parameter.
However, the realistic and
phenomenological interactions are quite different.
First of all, the realistic interactions have short-range cores,
which makes the application to many-body systems extremely difficult.
Also, they
have quite different interaction ranges.
Most of the realistic interactions have the energy minimum point around $r\sim1$~fm, where $r$ is the relative
distance between the nucleons.
In contrast, the ranges of the conventional interactions for the cluster studies
are much larger; 
for instance, the attractive term of the Volkov No.2 interaction has the range of 1.80~fm,
which creates the energy minimum point around $r\sim$1.3~fm for the even-parity state.
Therefore, the attractive effect has much longer ranges in the
phenomenological interactions.
It is quite interesting to investigate how the cluster structures appear with realistic interactions,
which have completely different ranges.

In this paper,
we discuss the cluster states of light nuclei starting with the realistic interaction.
The short-range correlation of the realistic interaction is treated by employing the damping factor.
Using an interaction transformed in this way, the $\alpha$-$\alpha$ energy curve and the scattering phase shift 
are calculated, which are compared with
the results of phenomenological interactions.
It is quite well known that
the phenomenological interaction 
should have proper density dependence in order
to satisfy the saturation property of nuclear systems.
The Tohsaki interaction, which has finite-range three-body terms~\cite{PhysRevC.49.1814,PhysRevC.94.064324,PTP.94.1019},
satisfies the saturation properties.
We employ this interaction as a phenomenological one, 
which gives a reasonable size and binding energy of the $\alpha$ cluster
in addition to the $\alpha$-$\alpha$ scattering phase shift.

Then the  interactions introduced here are
applied to the linear-chain states of $\alpha$ clusters.
There has been a long history of both theoretical and experimental studies 
for the possibility of $\alpha$ chain states~\cite{RevModPhys.90.035004}.  
Once the second $0^+$ state of $^{12}$C just above the three-$\alpha$ threshold energy 
has been assigned as a candidate, 
but now the state is regarded as an $\alpha$ gas state~\cite{PhysRev.101.254,Uegaki12C}. 
In return, not exactly linear but bent chain was suggested
around $E_x \sim10$~MeV region corresponding to 3~MeV above the three $\alpha$ threshold~\cite{PhysRevC.84.054308,NEFF2004357,PhysRevC.91.024315}. 
Furthermore, the possibility of four-$\alpha$ and 
six-$\alpha$ chain state was suggested~\cite{PhysRev.160.827,PhysRevLett.68.1295}.
It has been theoretically discussed that
the main decay path of the linear-chain states 
passes through  the bending motion,
and adding valence neutrons and/or
giving angular momentum to the system work to prevent this motion~\cite{Flocard-PTP.72.1000,PhysRevC.64.014301,BENDER2003390,PhysRevC.74.067304,MARUHN20101,PhysRevC.82.044301,PhysRevC.83.021303,PhysRevLett.107.112501,PhysRevC.90.054307,PhysRevC.92.011303,PhysRevLett.112.062501,PhysRevLett.115.022501,PhysRevC.95.044320,PhysRevC.97.054315,PhysRevC.98.044312,Ren2019}.
Not only cluster models, but recently various mean-field approaches are also successfully applied to 
discuss the stability of the linear-chain configurations.

This paper is organized as follows;
in Sec.~{\bf II}, the framework,
especially for the model wave function
and the Hamiltonian of the present model, is described.
In Sec.~{\bf III}, we introduce the damping factor for the short-range
part of the central interaction, and
the $\alpha$-$\alpha$ energy curve and scattering
phase shift are calculated and compared with
the results of a phenomenological interaction with
finite-range three-body terms ({\bf A.}). 
Also the linear-chain structure of three and four $\alpha$ clusters
in $^{12}$C and $^{16}$O
are calculated and compared with the results of the conventional approaches  ({\bf B.}).
The summary is presented in Sec.~{\bf IV}.

\section{Method}

\subsection{wave function (Brink model)}
Here we employ the conventional Brink model~\cite{Brink}
for the $\alpha$ cluster states.
In the Brink model, each single-particle wave function is described by a Gaussian,
\begin{equation}	
	\phi_{ij} = \left(  \frac{2\nu}{ \pi } \right)^{\frac{3}{4}} 
		\exp \left[-  \nu \left(\bm{r} - \bm{R}_i \right)^{2} \right] \chi_j, 
\label{spwf} 
\end{equation}
where the Gaussian center parameter $\bm{R}_i$ shows the expectation 
value of the position of the particle,
and $\chi_j$ is the spin-isospin wave function.
The size parameter $\nu$ is chosen to be 0.25~fm$^{-2}$,
which reproduces the observed radius of $^4$He.
Here, four single particle 
wave functions with different spin and isospin
sharing a common Gaussian center parameter
$\bm{R}_i$ correspond to an $\alpha$ cluster.
The Slater determinant of the Brink model is constructed from 
these single particle wave functions by antisymmetrizing them,
\begin{eqnarray}
\Phi_{SD}(\bm{R}_1, \bm{R}_2, \ldots, \bm{R}_N)
= &&  
{\cal A} \{ (\phi_{11}\phi_{12}\phi_{13}\phi_{14})  (\phi_{21}\phi_{22}\phi_{23}\phi_{24})   \nonumber \\
&& \ldots (\phi_{N1}\phi_{N2}\phi_{N3}\phi_{N4}) \}.
\label{SD} 
\end{eqnarray}
This is the case that we have $N$ $\alpha$ clusters and 
the mass number $A$ is given by $A=4N$.
This assumption of common Gaussian center parameter for
four nucleons removes the contribution
of the non-central interactions, spin-orbit and tensor components,
after the antisymmetrization.

For  $^8$Be we introduce two $\alpha$ clusters, and two Gaussian 
center parameters are introduced
as $\bm{R}_1  = d \bm{e}_z/2$ and $\bm{R}_2 -d \bm{e}_z/2$
with the relative distance of $d$,
where $\bm{e}_z$ is the unit vector for the $z$ direction.
For the ground state of $^{16}$O, the Gaussian center parameters are introduced to have
a tetrahedron shape with a fixed relative distance $d$.
In both cases,
the Slater determinants 
with different $d$ values are superposed
based on the generator coordinate method (GCM)~\cite{Brink}.

Also, the linear chain configurations of three and four $\alpha$ clusters
in $^{12}$C and $^{16}$O can be calculated by assuming one-dimensional configurations
for the $\{ \bm{R}_i \}$ values. The distances between the $\alpha$ clusters
are randomly generated, and these Slater determinants are superposed
based on the GCM.

The Slater determinants are projected to the eigen states of the angular momenta
by numerical integration,
\begin{equation}
\Phi^J_{MK} = {2J+1 \over 8\pi^2} \int d\Omega {D_{MK}^J}^*R(\Omega) \Phi_{SD},
\end{equation}
where ${D_{MK}^J}$ is Wigner D-function 
and $R(\Omega)$ is the rotation operator
for the spatial and spin parts of the wave function.
This integration over the Euler angle $\Omega$ is numerically performed.

\subsection{Hamiltonian}  
The Hamiltonian used in the present calculation is
\begin{align}
\hat{H}&=\sum_{i}^A
\hat{T}_i-\hat{T}_\mathrm{G} \nonumber \\
&
+\sum_{i<j}^A 
\left[
\hat{V}_\mathrm{\cent}(i,j)+\hat{V}_\mathrm{Coulomb}(i,j)\right],
\end{align}
where $\hat{T}_i$ is the kinetic energy operator of $i$th nucleon,
and the total kinetic energy operator for the cm motion ($\hat{T}_\mathrm{G}$)
is subtracted.
The two-body interaction consists of only the central interaction
($\hat{V}_\mathrm{\cent}$) and 
Coulomb interaction ($\hat{V}_\mathrm{Coulomb}$),
since the effect of the non-central interactions disappear 
in the $\alpha$ cluster model.
This  above Hamiltonian is relevant for the G3RS interaction, which is 
a realistic interaction.
However, 
to reduce the height of the short-range core part. For the region of $r \leq r_0$,
the factor $F(r)$
is multiplied to the central part of the G3RS interaction.
Also, we examine the effect of the finite-range three-body interaction.

We compare the results obtained with a phenomenological interaction,
Tohsaki F1 interaction~\cite{PhysRevC.49.1814,PhysRevC.94.064324,PTP.94.1019}.
This interaction also contains finite-range three-body terms.

\section{Results}

\subsection{introduction of the damping factor}

For the central interaction $\hat{V}_\mathrm{\cent}$, 
we use the G3RS interaction~\cite{PTP.39.91}, which is the realistic interaction
designed to reproduce the scattering phase shift of the nucleon-nucleon scattering
up to around 600~MeV.
The central part of G3RS has three-range Gaussian form as
\begin{align}
\widehat{V}_\mathrm{c}=&
\widehat{P}_{ij}(^3E)\sum_{n=1}^3V_{\mathrm{c},n}^{^3E}\exp\left(-\frac{r_{ij}^2}{\eta_{\mathrm{c},n}^2}\right)\nonumber\\
&+\widehat{P}_{ij}(^1E)\sum_{n=1}^3V_{\mathrm{c},n}^{^1E}\exp\left(-\frac{r_{ij}^2}{\eta_{\mathrm{c},n}^2}\right)\nonumber\\
&+\widehat{P}_{ij}(^3O)\sum_{n=1}^3V_{\mathrm{c},n}^{^3O}\exp\left(-\frac{r_{ij}^2}{\eta_{\mathrm{c},n}^2}\right)\nonumber\\
&+\widehat{P}_{ij}(^1O)\sum_{n=1}^3V_{\mathrm{c},n}^{^1O}\exp\left(-\frac{r_{ij}^2}{\eta_{\mathrm{c},n}^2}\right), 
\end{align}
where $\widehat{P}_{ij}(^{1,3}E)$ and  $\widehat{P}_{ij}(^{1,3}O)$ are the projection operators to the 
$^{1,3}E$ (singlet-even, triplet-even) and $^{1,3}O$  (singlet-even, triplet-odd) states, respectively. 
The parameter set of ``case 1'' in Ref.~\cite{PTP.39.91} is listed 
in Table \ref{g3rs-parameter}. 

\begin{table}[!thbp]
\caption{The parameter sets for the central part of the G3RS interaction. 
The parameter set ``case 1'' in Ref.~\cite{PTP.39.91} is adopted.}
\label{g3rs-parameter}
\centering
\begin{tabular}{lrrr}
\hline \hline
$n$	&$	1	$&$	2	$&$	3	$\\
\hline
$\eta_{\textrm{c},n}$ (fm)	& $2.5$ & $0.942$ & $0.447$ \\
$V_{\textrm{c},n}^{^3E}$ (MeV)	& $-5$ & $-210$ & $2000$ \\
$V_{\textrm{c},n}^{^1E}$ (MeV)	& $-5$ & $-270$ & $2000$ \\
$V_{\textrm{c},n}^{^3O}$ (MeV)	& $1.6667$ & $-50$ & $2500$ \\
$V_{\textrm{c},n}^{^1O}$ (MeV)	& $10$ & $50$& $2000$ \\
\hline\hline
\end{tabular}
\end{table}

\begin{figure}[htbp]
	\centering
	\includegraphics[width=6.5cm]{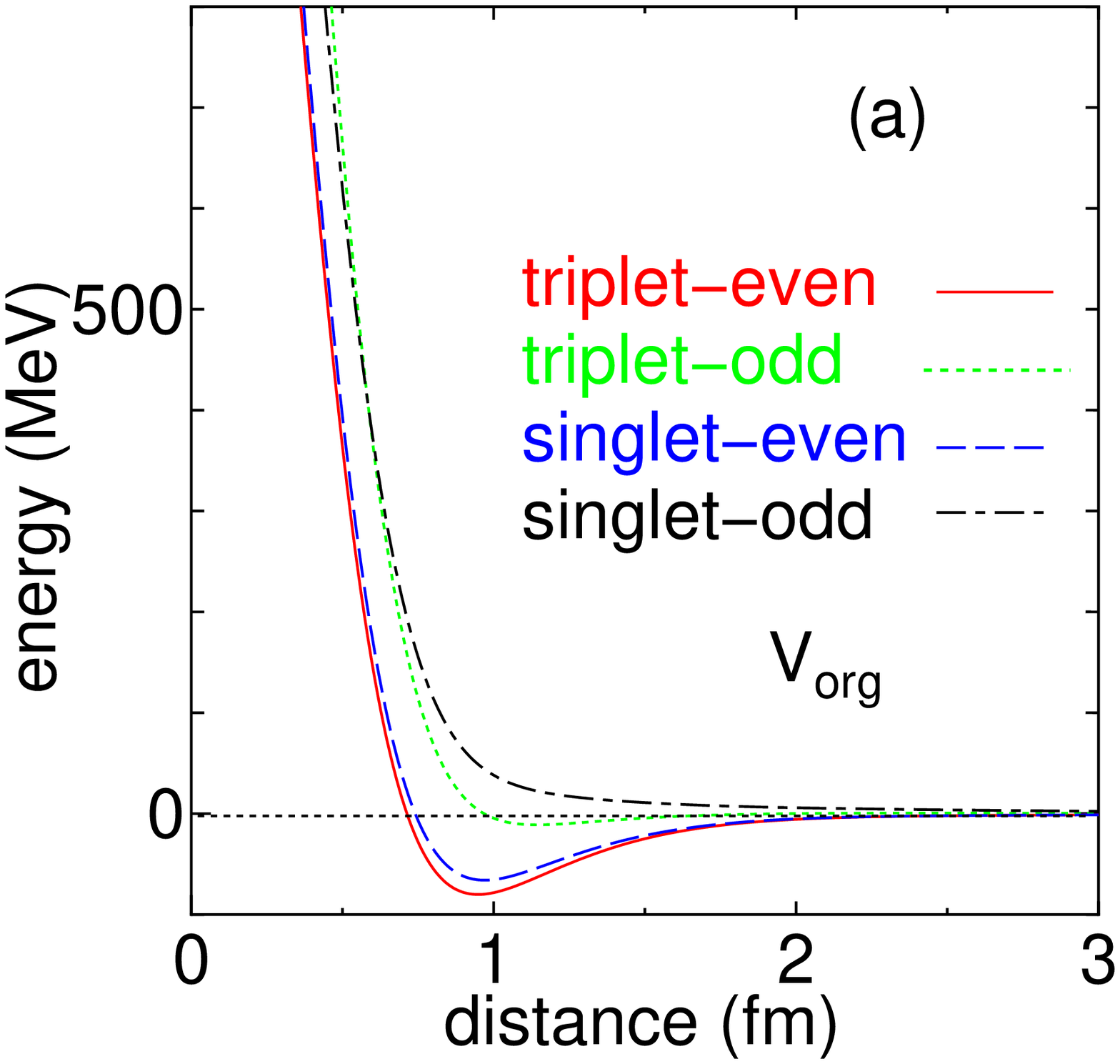}
	\includegraphics[width=6.5cm]{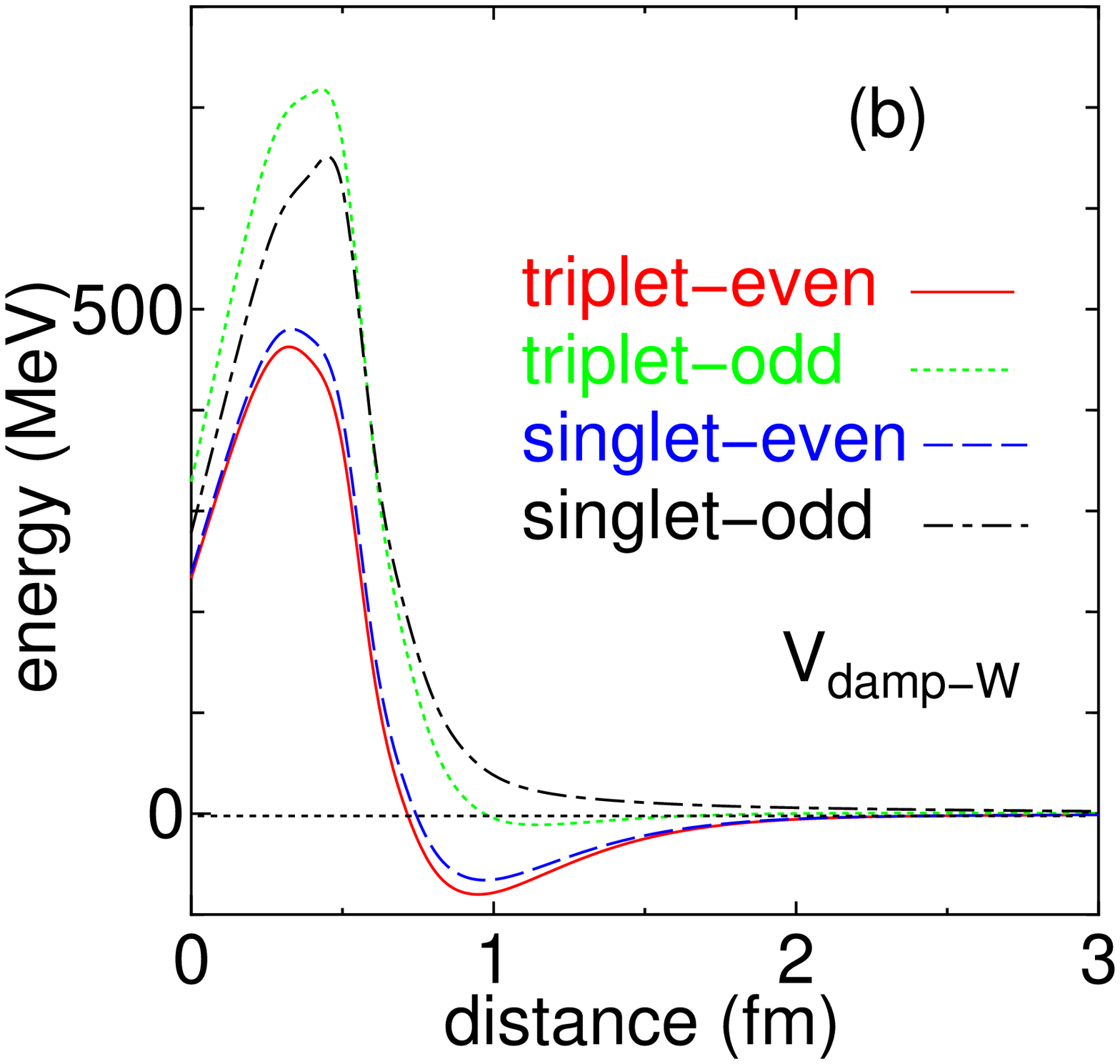}  
	\includegraphics[width=6.5cm]{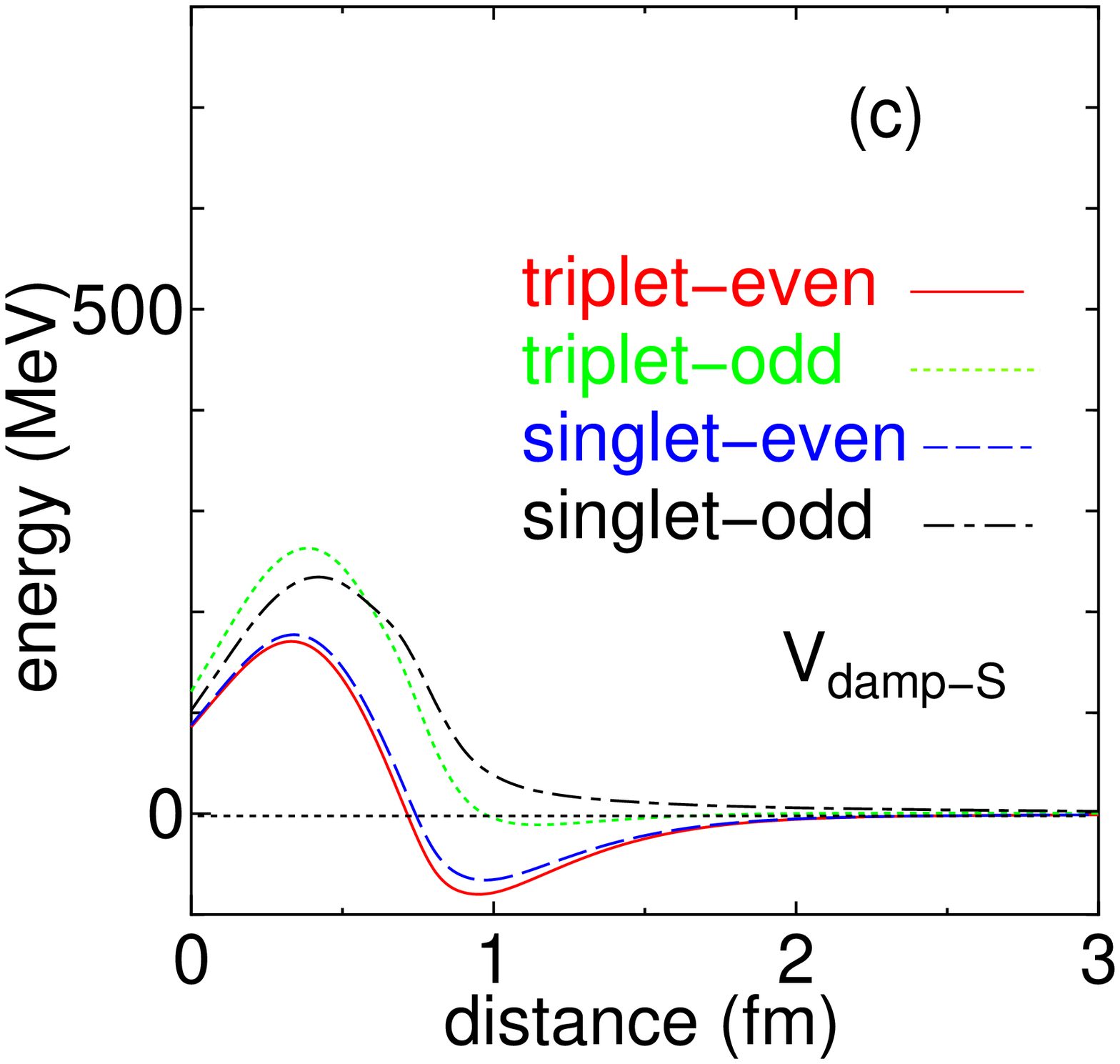} 
	\caption{
The spatial part of the G3RS interaction as functions of the nucleon-nucleon distance $r$
(``case 1'' in Ref.~\cite{PTP.39.91}).
(a): $V_{org}$, the original form, 
(b): $V_{damp-W}$, with weaker damping factor, $r_0 = 0.5$~fm and $m=2$ in Eq.~(\ref{eq:damp}),
and (c): $V_{damp-S}$, with stronger damping factor, $r_0 = 0.75$~fm and $m=3$ in Eq.~(\ref{eq:damp}).
}
\label{g3rs-r}
\end{figure}

The spatial part of the G3RS interaction as functions of the nucleon-nucleon distance $r$
is shown in Fig.~\ref{g3rs-r}.
Here, Fig.~\ref{g3rs-r} (a) is the original form of the G3RS interaction,
which is called $V_{org}$.
This is the realistic interaction and has a short-range core part.
For the use of the calculation of the many-body system, here we introduce a damping factor
to reduce the height of the short-range core part (around 2500~MeV). 
This is equivalent to the regularization in 
the momentum space to suppress the high-momentum contribution 
(see, for instance Ref.~\cite{MACHLEIDT20111}
for the regularization in the chiral effective field theory).
For the region of $r \leq r_0$,
the factor $F(r)$
is multiplied to the central part of the G3RS interaction;
\begin{equation}
F(r) = 1/(\exp[1-r/r_0])^{m}.
\label{eq:damp}
\end{equation}
This damping factor is unity in the region of  $r\geq r_0$ but
start changing at $r = r_0$ and converges to $e^{-m}$ at $r = 0$.
In this article, we compare two parameter sets; weaker damping factor and stronger one.
For the weaker damping factor, the values of  $r_0 = 0.5$~fm and $m = 2$ are 
adopted. The resultant G3RS interaction is called $V_{damp-W}$ and the form is shown in
Fig~\ref{g3rs-r}~(b).
This is the G3RS interaction after multiplying the damping factor 
in the region of $r  \leq 0.5$~fm, and the short-range core is reduced.
We also introduce stronger damping factor,
which 
has the parameters of
$r_0 = 0.75$~fm and $m = 3$ and
is called $V_{damp-S}$.
Figure~\ref{g3rs-r}~(c) shows the shape after multiplying
the damping factor.
As we can see, when two nucleons get closer,
the damping starts already at $r = 0.75$~fm and
the inner part is drastically reduced.

\begin{figure}[htbp]
	\centering
	\includegraphics[width=6.5cm]{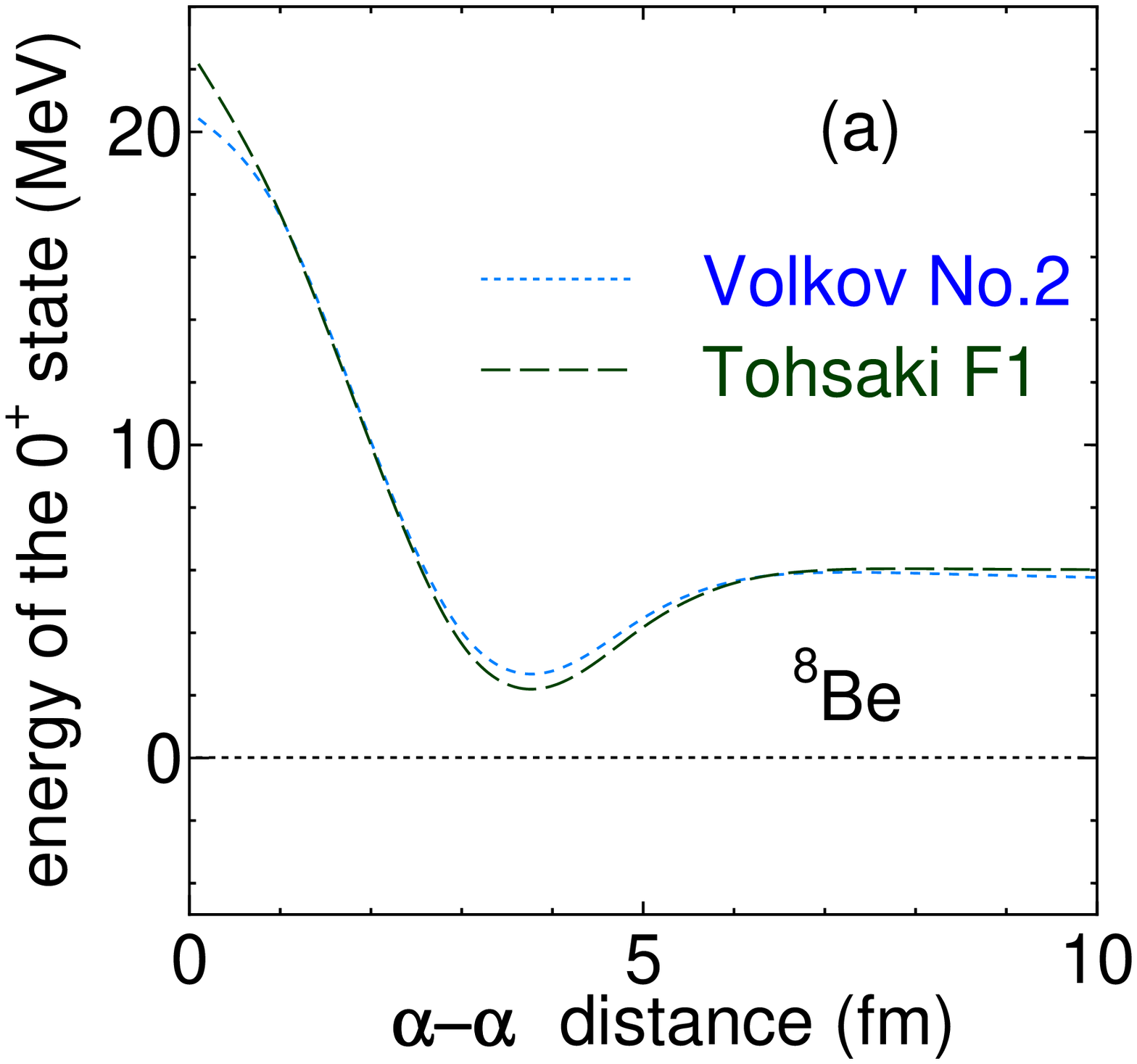} 
	\includegraphics[width=6.5cm]{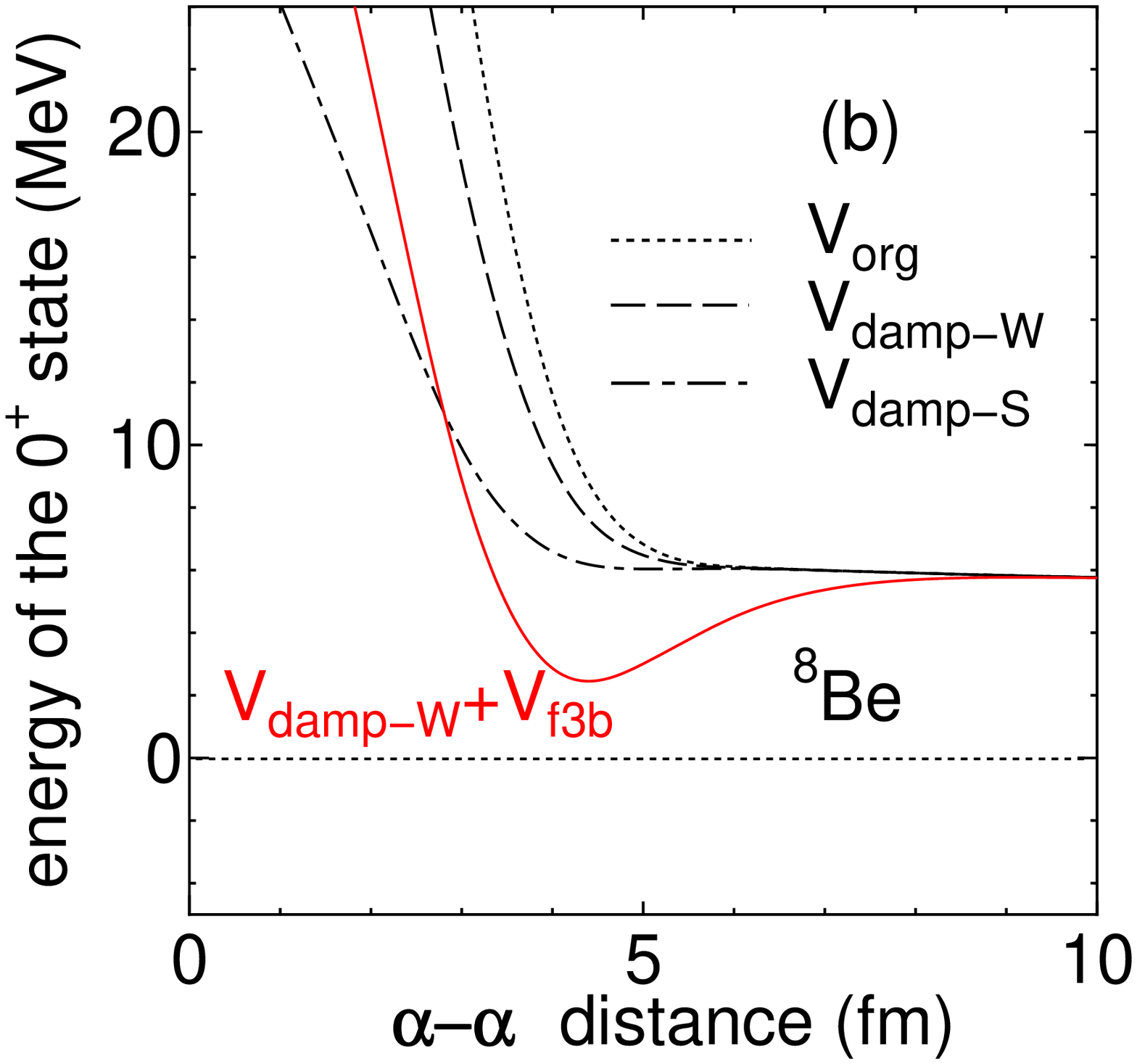} 
	\caption{
The $0^+$ energy curves of  two $\alpha$ clusters ($^8$Be). 
The energy is measured from the two $\alpha$ threshold.
The horizontal axis shows the distance between the two $\alpha$ clusters.
(a):  the  dotted and dashed lines are obtained with the Volkov No.2 interaction
with the Majorana exchange parameter of $M=0.60$ and  Tohsaki F1 interaction,
respectively.
(b): the dotted, dashed, and dot-dashed lines are results of
$V_{org}$,
$V_{damp-W}$,
and $V_{damp-S}$, respectively.
The solid line corresponds to the result 
by $V_{damp-W}$ with a finite-range three-body term ($V_{damp-W}+V_{f3b}$).
}
\label{alpha-alpha}
\end{figure}

Using these interactions, 
the $0^+$ energy curves of  two $\alpha$ clusters ($^8$Be)
are calculated and shown in Fig.~\ref{alpha-alpha}.
At first, we show the energy curves of the phenomenological interactions,
which are known to well reproduce the $\alpha$-$\alpha$ scattering phase shift.
The dotted line and dashed line in Fig.~\ref{alpha-alpha}~(a)
are results of the Volkov No.2 interaction~\cite{VOLKOV196533} with the Majorana exchange parameter
of $M = 0.6$ and Tohsaki F1 interaction~\cite{PhysRevC.49.1814}, respectively.
The horizontal axis shows the distance between the two $\alpha$ clusters,
and the energy is measured from the two $\alpha$ threshold.
These two lines are very similar, and both of them have the energy minimum points
around the $\alpha$-$\alpha$ distance of 4~fm.
This is coming from the long-range nature of these phenomenological interactions.
The energy does not become zero even at the
large relative distances owing to the zero-point kinetic energy
($\hbar \omega/4 \sim 5$~MeV),
which is the kinetic energy for fixing the relative distance.

Next, the energy curves of $\alpha$-$\alpha$ calculated with the G3RS interaction
are shown in Fig.~\ref{alpha-alpha}~(b).
The energy of each $\alpha$ cluster indeed depends on the interactions,
but again here the energy is measured from the two-$\alpha$ threshold.
The dotted line is the result of the original G3RS interaction,
$V_{org}$ (Fig.~\ref{g3rs-r}~(a)).
Due to the repulsive effect at short relative distances,
the energy minimum point around the relative distance of 4~fm
found in Fig.~\ref{alpha-alpha}~(a)
does not appear. 
This energy minimum point cannot be obtained
even if we 
reduce the short-range repulsion
by introducing
the damping factor for the nucleon-nucleon
interaction, as in the dashed line, where $V_{damp-W}$ is adopted;
the energy around the $\alpha$-$\alpha$ distance of 4~fm is constant 
around 6~MeV.
Rather surprisingly, this situation does not change even if we adopted a much stronger damping factor.
The dash-dotted line shows the result of $V_{damp-S}$, and even if the repulsive effect is reduced
at short $\alpha$-$\alpha$ distances, the energy around 4~fm is still constant around 6~MeV.
Experimentally, the ground $0^+$ state of $^8$Be is
located at 0.09184~MeV, and the $\alpha$-$\alpha$ energy within the 
region inside the Coulomb barrier 
is essential for the reproduction of this value.
If we superpose Slater determinants 
with different $\alpha$-$\alpha$ distances (1, 2, 3,....10~fm)
based on the GCM,
the obtained energy of the ground $0^+$ state
does decrease by a few MeV.
Nevertheless, 
it is still above the $\alpha$-$\alpha$ threshold by 1-2~MeV
in the cases of $V_{org}$, $V_{damp-W}$, and even $V_{damp-S}$.
Without the reproduction of this ground state energy, we cannot discuss the scattering 
phase shift, which is quite sensitive to the energy of the resonance state from the threshold.
Therefore, we need an additional effect 
other than reducing  
the short-range repulsion by introducing the damping factor, 
which contributes to the 
lowering the energy around the $\alpha$-$\alpha$
distance of 4~fm.

One may consider that the missing of the tensor interaction 
due to the assumption of $(0s)^4$ configuration for each $\alpha$ cluster
is the origin of the shortage of attractive effect.
However,  as we discussed based on the AQCM-T~\cite{PTEP-Itagaki-ptz046,PhysRevC.98.054306},
the attractive effect of the tensor interaction is very strong inside each $^4$He
rather than between the two $^4$He nuclei, when 
two $^4$He nuclei are far with each other. This tensor effect
is blocked  when two $^4$He approach, which works repulsively.
One of the other mechanisms, which contributes to the 
lowering of the energy around the $\alpha$-$\alpha$
distance of 4~fm,
is three-body nuclear interaction.
Here we adopt a finite-range three-body interaction $V_{f3b}$,
\begin{equation}
V_{f3b} = {1 \over 6} \sum_{i \neq j, j \neq k, i \neq k}  V^{(3)}_{ijk},
\end{equation}
where,
\begin{eqnarray}
V^{(3)}_{ijk} = V^{(3)} && \exp[-\mu (\vec r_i - \vec r_j )^2 - \mu (\vec r_i - \vec r_k)^2    ] \nonumber \\
\times &&  (W + M P^r_{ij}) (W + M P^r_{ik}).
\label{three-body}
\end{eqnarray}
The parameters involved in Eq.~(\ref{three-body}) 
are fixed phenomenologically in order to account for the 2-$\alpha$ system, as well as the properties of $^{16}$O.
The strength $V^{(3)}$ and the range $\mu$ are set to $-9$~MeV and 0.1~fm$^{-2}$,
respectively. The strength is small, but it has a very long range.
The Majorana exchange parameter $M$ is set to 0.645 and $W = 1-M$.
The operator $P^r_{ij}$ exchanges the spatial part of the wave functions 
of the interacting $i$-th and $j$-th nucleons.
The introduction of this Majorana term is needed for the reproduction 
of the binding energy of $^{16}$O from the four-$\alpha$ threshold,
which will be discussed shortly.
With this phenomenological three-body nucleon-nucleon interaction,
we do not need to drastically reduce the short-range repulsion of the two-body interaction,
thus we adopt weaker damping factor, $V_{damp-W}$ (extremely strong damping factor gives poor reproduction of the $\alpha$-$\alpha$ scattering phase shift).
The $\alpha$-$\alpha$ energy curve calculated with 
$V_{damp-W}+V_{f3b}$ is shown as the solid line in Fig.~\ref{alpha-alpha}~(b).
The energy minimum point appears around the $\alpha$-$\alpha$ distance of 4~fm,
and the shape is very similar,
apart from the short-distance region, 
to the results of the phenomenological interactions 
shown in Fig.~\ref{alpha-alpha}~(a),
which reproduce the $\alpha$-$\alpha$ scattering phase shift.

\begin{figure}[htbp]
	\centering
     \includegraphics[width=6.5cm]{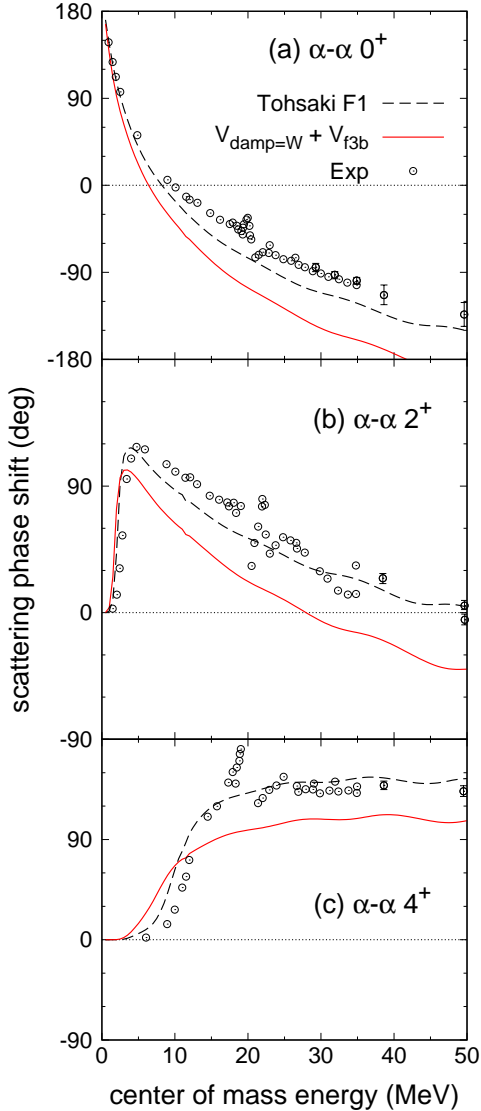} 
	\caption{
The $\alpha$-$\alpha$ scattering phase shifts
as a function of the center of mass energy between two $\alpha$'s,
(a): $0^+$, (b): $2^+$, and (c): $4^+$.
The  solid lines show the results of $V_{damp-W}+V_{f3b}$,
and the dashed lines are those
of Tohsaki F1 interaction~\cite{PhysRevC.49.1814}.
The experimental data (open circles) are taken from Ref.~\cite{10.1143/PTP.53.677}, 
where the measured data were complied from the original works~\cite{PhysRev.104.123,PhysRev.104.135,PhysRev.109.850,PhysRev.117.525,MiyakeBICR1961,PhysRev.129.2252,PhysRevLett.29.1331}.
}
\label{aa-ps}
\end{figure}

The $\alpha$-$\alpha$ scattering phase shift  ($^8$Be)
is calculated and
shown in Fig.~\ref{aa-ps} ((a): $0^+$, (b): $2^+$, and (c): $4^+$)
as a function of the center of mass energy of two $\alpha$'s.
This is calculated using the
Kohn-Hulth\'{e}n variational principle combined with the GCM~\cite{Mito.PTP.56.583,Kamimura.PTPS.62.236}.
The  solid lines show the results of $V_{damp-W}+V_{f3b}$,
and the dashed lines are obtained with the Tohsaki F1 interaction,
which is designed to reproduce the $\alpha$-$\alpha$ scattering phase shift.
The experimental data (open circles) are taken from Ref.~\cite{10.1143/PTP.53.677}, 
where the measured data were complied from the original works~\cite{PhysRev.104.123,PhysRev.104.135,PhysRev.109.850,PhysRev.117.525,MiyakeBICR1961,PhysRev.129.2252,PhysRevLett.29.1331}.
The reproduction of the data is not perfect and the repulsive effect is a bit stronger in the cases of 
$V_{damp-W}+V_{f3b}$, but the basic trend of the scattering phase shift can be satisfied.

\begin{figure}[htbp]
	\centering
	\includegraphics[width=6.5cm]{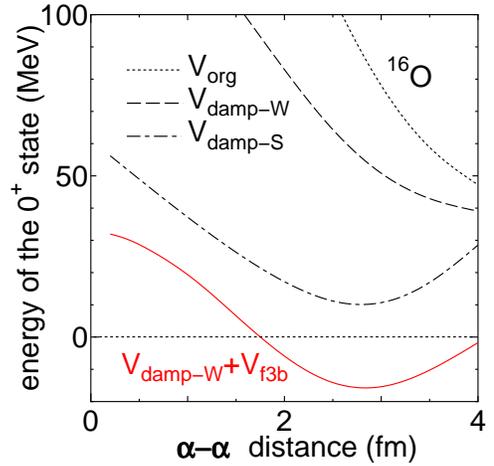} 
	\caption{
The $0^+$ energy curves of $^{16}$O with the tetrahedron configuration of four $\alpha$ clusters
as a functions of the relative $\alpha$-$\alpha$ distance $d$. 
The energy is measured from the four $\alpha$ threshold.
The lines are the same as in Fig.~\ref{alpha-alpha}.
}
\label{o-tetra}
\end{figure}

Figure~\ref{o-tetra} shows the $0^+$ energy of $^{16}$O.
Here tetrahedron configuration of four $\alpha$ clusters is assumed, 
and the horizontal axis shows the distance between the two $\alpha$ clusters.
The dotted, dashed, and dot-dashed lines 
are respectively associated with
$V_{org}$,
$V_{damp-W}$,
and $V_{damp-S}$.
The solid line is the result by $V_{damp-W}$ with finite-range three-body 
interaction ($V_{damp-W}+V_{f3b}$),
and the energy minimum point appears at the $\alpha$-$\alpha$
distance of 2.5-3.0~fm.
Experimentally, the ground state
of $^{16}$O is lower than
the four-$\alpha$ threshold energy by 14.4~MeV,
and the solid line is close to this value.
We superpose the Slater determinants with different $\alpha$-$\alpha$
distances based on the GCM, and the ground $0^+$ state is obtained at
$-16.2$~MeV. The matter radius is obtained as 2.52~fm corresponding
to the charge radius of 2.65~fm (experimentally 2.69~fm).

\subsection{linear-chain states}

\begin{figure}[htbp]
	\centering
      \includegraphics[width=6.5cm]{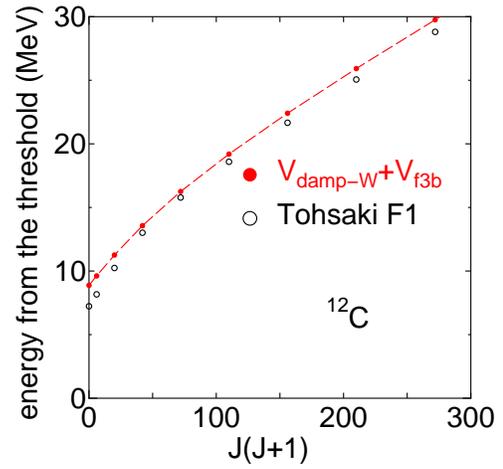} 
	\caption{
The rotational band structure of the linear-chain structure of the three $\alpha$ clusters ($^{12}$C). 
The energy is measured from the three-$\alpha$ threshold.
The horizontal axis is $J(J+1)$, where $J$ is the angular momentum of the system.
The solid circles show the result for $V_{damp-W}+V_{f3b}$.
The open circles are the result obtained with a phenomenological interaction, Tohsaki F1~\cite{PhysRevC.49.1814}.
}
\label{c-lin}
\end{figure}

The interaction introduced in the last subsection can be applied to the linear-chain configurations of 
three and four $\alpha$ clusters in $^{12}$C and $^{16}$O.
The rotational band structure of the linear-chain structure of the three $\alpha$ clusters ($^{12}$C)
is shown in Fig.~\ref{c-lin}. The Slater determinants with different distances between $\alpha$ clusters
within one-dimensional configurations
are randomly generated and superposed based on the GCM.
The energy is measured from the three $\alpha$ threshold.
The horizontal axis is $J(J+1)$, where $J$ is the angular momentum of the system.
The  solid circles show the result of $V_{damp-W}+V_{f3b}$
and open circles 
correspond to that of
 Tohsaki F1.
The band head energy is 8.9~MeV for $V_{damp-W}+V_{f3b}$  (solid circle),
and Tohsaki interaction gives 7.2~MeV (open circle).
Experimentally
the three-$\alpha$ threshold energy corresponds to
$E_x=7.4$~MeV,
thus these band head energies correspond to $E_x = 16-17$~MeV from the ground state.
The band head energy 
was estimated as $E_x = 16.6$~MeV (measured from the ground state)
in the covariant density functional approach~\cite{PhysRevLett.115.022501} and 
$E_x = 10.1$~MeV (measured from the three-$\alpha$ threshold) in a modern cluster model~\cite{PhysRevLett.112.062501}.
Our results of both $V_{damp-W}+V_{f3b}$ and  Tohsaki F1  
almost agree with these.
The slope of the rotational band ($\hbar^2/2I$, $I$ the moment of inertia)
is obtained as 
102~keV and 119~keV
for the solid and open circles, respectively,
if we just average between the $0^+$ and $8^+$ energies.
However in the high-spin states,
where the centrifugal force is strong,
the $\alpha$-$\alpha$ distance gets larger,
which creates a decrease of the slope.
This is in $^{14}$C and not $^{12}$C, but the slope of
$\hbar^2 / 2I \sim$ 120 keV was reported in Ref.~\cite{PhysRevC.90.054324}.

\begin{figure}[htbp]
	\centering
	\includegraphics[width=6.5cm]{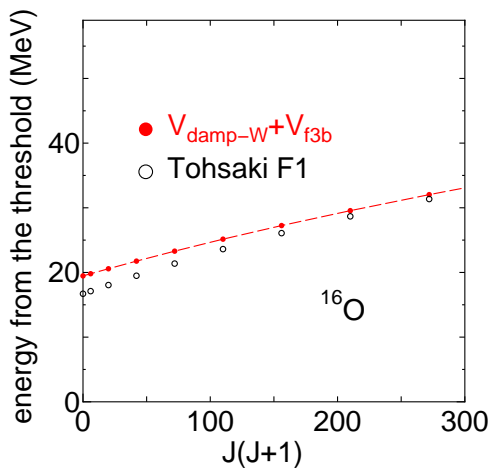} 
	\caption{
The rotational band structure of the linear-chain structure of the four $\alpha$ clusters ($^{16}$O). 
The energy is measured from the four $\alpha$ threshold.
The circles are the same as in Fig.~\ref{c-lin}.
}
\label{o-lin}
\end{figure}

The rotational band structure of the linear-chain structure of the four $\alpha$ clusters ($^{16}$O)
is shown in Fig.~\ref{o-lin}. Again the Slater determinants with different distances between $\alpha$ clusters
are randomly generated  and superposed based on the GCM,
and the energy is measured from the four $\alpha$ threshold.
The horizontal axis is $J(J+1)$, where $J$ is the angular momentum of the system.
The  solid circles show the result of $V_{damp-W}+V_{f3b}$
and the open circles are associated with Tohsaki F1.
The band head energy measured from the four-$\alpha$ threshold energies
are 19.4~MeV and 16.7~MeV 
for the solid and open circles, respectively.
If we measure the energy from the ground state, 
$V_{damp-W}+V_{f3b}$ 
(solid circle) give  the band head energy of 35.6~MeV.
The slope of the rotational band 
again decreases in the high-spin states 
due to the large centrifugal force 
(an increase of the $\alpha$-$\alpha$ distance).
The solid circles give the value of the slope as 0.051~MeV 
between the $0^+$ and $10^+$ states
(open circles give 0.063~MeV).
These interactions give similar results.

These days, the linear-chain states of four $\alpha$ clusters are discussed
also with various density functional theories~\cite{PhysRevLett.107.112501,PhysRevC.90.054307}.
Among them, 
nonrelativistic Hartree-Fock approaches are adopted in
Refs.~\cite{Flocard-PTP.72.1000,BENDER2003390,PhysRevLett.107.112501},
and the band head energy of $\sim40$~MeV from the ground state,
which agrees with the results of the present study,
was reported~\cite{PhysRevLett.107.112501}.
The slope of 
$\hbar^2/2I = 0.06-0.08$~MeV was suggested there.
A covariant framework predicted
the band head energy of $\sim30$~MeV and the slope of 
$\hbar^2/2I = 0.11$~MeV~\cite{PhysRevC.90.054307}.
Similar values are obtained by the modern cluster model studies~\cite{PhysRevLett.112.062501,PhysRevC.95.044320}.

\section{Conclusion}
 We investigated the cluster states of light nuclei starting with the realistic interaction, G3RS.
The short-range correlation of the realistic interaction was treated by employing 
a damping factor.
Using an interaction transformed in such a way, the $\alpha$-$\alpha$ energy curve and the scattering  
phase shift were calculated.
Although the original G3RS interaction shows too repulsive features,
the scattering phase shift is reasonably reproduced after introducing the damping factor.
The inclusion of a finite-rang three-body interaction gives the energy minimum point of 
the $\alpha$-$\alpha$ energy curve around the relative distance of 4~fm
as in the case of the phenomenological interaction.
This allows the reproduction of the ground-state energy, which is essential 
for describing the scattering phase shift,
but this energy minimum point does not appear without the three-body effect
even with a very strong damping factor 
for the short-range repulsion of the two-body interaction.
With this three-body term, it is capable of reproducing
the binding energy of four $\alpha$ clusters from the threshold in $^{16}$O.

The rotational band structure of the linear-chain structure of the three $\alpha$ clusters ($^{12}$C)
can be calculated, and
the band head energy was obtained at 8.9~MeV
from the three-$\alpha$ threshold energy in the case of G3RS interaction
with the three-body interaction.
This result well agrees with  those by other conventional approaches.
Similarly, the linear-chain structure of the four $\alpha$ clusters ($^{16}$O)
can be calculated and the band head energy was calculated at $\sim$36~MeV.
These days, the linear-chain states of four $\alpha$ clusters are discussed
also with various density functional theories,
which predicted the bad head energies of $30\sim40$~MeV 
from the ground state, in agreement with the result of the present study.
The slope of the rotational band structure of the G3RS interaction also
almost agrees with the phenomenological cluster models
and density functional theories.

As future works,
using the interaction proposed here, light neutron-rich nuclei can be calculated.
The strength of the three-body interaction should be carefully determined,
and introducing Heisenberg and Bartlet exchange terms is the possible way
to keep the consistency with the present results.
Also, deriving an effective interaction consisting of the two- 
and three-body terms for the cluster model in a more microscopic way is quite an important task.
We are working along this line based on the chiral effective field theory.

\begin{acknowledgments}
We thank Professor M.~Kamimura
for the calculation of the scattering phase shift.
The numerical calculation has been performed using the computer facility of 
Yukawa Institute for Theoretical Physics,
Kyoto University and Research Center for Nuclear Physics, Osaka University.
\end{acknowledgments}

\bibliography{biblio_ni.bib}

\begin{thebibliography}{61}%
\makeatletter
\providecommand \@ifxundefined [1]{%
 \@ifx{#1\undefined}
}%
\providecommand \@ifnum [1]{%
 \ifnum #1\expandafter \@firstoftwo
 \else \expandafter \@secondoftwo
 \fi
}%
\providecommand \@ifx [1]{%
 \ifx #1\expandafter \@firstoftwo
 \else \expandafter \@secondoftwo
 \fi
}%
\providecommand \natexlab [1]{#1}%
\providecommand \enquote  [1]{``#1''}%
\providecommand \bibnamefont  [1]{#1}%
\providecommand \bibfnamefont [1]{#1}%
\providecommand \citenamefont [1]{#1}%
\providecommand \href@noop [0]{\@secondoftwo}%
\providecommand \href [0]{\begingroup \@sanitize@url \@href}%
\providecommand \@href[1]{\@@startlink{#1}\@@href}%
\providecommand \@@href[1]{\endgroup#1\@@endlink}%
\providecommand \@sanitize@url [0]{\catcode `\\12\catcode `\$12\catcode
  `\&12\catcode `\#12\catcode `\^12\catcode `\_12\catcode `\%12\relax}%
\providecommand \@@startlink[1]{}%
\providecommand \@@endlink[0]{}%
\providecommand \url  [0]{\begingroup\@sanitize@url \@url }%
\providecommand \@url [1]{\endgroup\@href {#1}{\urlprefix }}%
\providecommand \urlprefix  [0]{URL }%
\providecommand \Eprint [0]{\href }%
\providecommand \doibase [0]{http://dx.doi.org/}%
\providecommand \selectlanguage [0]{\@gobble}%
\providecommand \bibinfo  [0]{\@secondoftwo}%
\providecommand \bibfield  [0]{\@secondoftwo}%
\providecommand \translation [1]{[#1]}%
\providecommand \BibitemOpen [0]{}%
\providecommand \bibitemStop [0]{}%
\providecommand \bibitemNoStop [0]{.\EOS\space}%
\providecommand \EOS [0]{\spacefactor3000\relax}%
\providecommand \BibitemShut  [1]{\csname bibitem#1\endcsname}%
\let\auto@bib@innerbib\@empty
\bibitem [{\citenamefont {Brink}(1966)}]{Brink}%
  \BibitemOpen
  \bibfield  {author} {\bibinfo {author} {\bibfnamefont {D.~M.}\ \bibnamefont
  {Brink}},\ }\href@noop {} {\bibfield  {journal} {\bibinfo  {journal} {Proc.
  Int. School Phys.``Enrico Fermi"}\ }\textbf {\bibinfo {volume} {XXXVI}},\
  \bibinfo {pages} {247} (\bibinfo {year} {1966})}\BibitemShut {NoStop}%
\bibitem [{\citenamefont {Fujiwara}\ \emph {et~al.}(1980)\citenamefont
  {Fujiwara}, \citenamefont {Horiuchi}, \citenamefont {Ikeda}, \citenamefont
  {Kamimura}, \citenamefont {Kat\ifmmode~\bar{o}\else \={o}\fi{}},
  \citenamefont {Suzuki},\ and\ \citenamefont {Uegaki}}]{PTPS.68.29}%
  \BibitemOpen
  \bibfield  {author} {\bibinfo {author} {\bibfnamefont {Y.}~\bibnamefont
  {Fujiwara}}, \bibinfo {author} {\bibfnamefont {H.}~\bibnamefont {Horiuchi}},
  \bibinfo {author} {\bibfnamefont {K.}~\bibnamefont {Ikeda}}, \bibinfo
  {author} {\bibfnamefont {M.}~\bibnamefont {Kamimura}}, \bibinfo {author}
  {\bibfnamefont {K.}~\bibnamefont {Kat\ifmmode~\bar{o}\else \={o}\fi{}}},
  \bibinfo {author} {\bibfnamefont {Y.}~\bibnamefont {Suzuki}}, \ and\ \bibinfo
  {author} {\bibfnamefont {E.}~\bibnamefont {Uegaki}},\ }\href@noop {}
  {\bibfield  {journal} {\bibinfo  {journal} {Progress of Theoretical Physics
  Supplement}\ }\textbf {\bibinfo {volume} {68}},\ \bibinfo {pages} {29}
  (\bibinfo {year} {1980})}\BibitemShut {NoStop}%
\bibitem [{\citenamefont {Hoyle}(1954)}]{Hoyle}%
  \BibitemOpen
  \bibfield  {author} {\bibinfo {author} {\bibfnamefont {F.}~\bibnamefont
  {Hoyle}},\ }\href@noop {} {\bibfield  {journal} {\bibinfo  {journal}
  {Astrophys. J. Suppl.}\ ,\ \bibinfo {pages} {121}} (\bibinfo {year}
  {1954})}\BibitemShut {NoStop}%
\bibitem [{\citenamefont {Uegaki}\ \emph {et~al.}(1977)\citenamefont {Uegaki},
  \citenamefont {Okabe}, \citenamefont {Abe},\ and\ \citenamefont
  {Tanaka}}]{Uegaki12C}%
  \BibitemOpen
  \bibfield  {author} {\bibinfo {author} {\bibfnamefont {E.}~\bibnamefont
  {Uegaki}}, \bibinfo {author} {\bibfnamefont {S.}~\bibnamefont {Okabe}},
  \bibinfo {author} {\bibfnamefont {Y.}~\bibnamefont {Abe}}, \ and\ \bibinfo
  {author} {\bibfnamefont {H.}~\bibnamefont {Tanaka}},\ }\href@noop {}
  {\bibfield  {journal} {\bibinfo  {journal} {Progress of Theoretical Physics}\
  }\textbf {\bibinfo {volume} {57}},\ \bibinfo {pages} {1262} (\bibinfo {year}
  {1977})}\BibitemShut {NoStop}%
\bibitem [{\citenamefont {Tohsaki}\ \emph {et~al.}(2001)\citenamefont
  {Tohsaki}, \citenamefont {Horiuchi}, \citenamefont {Schuck},\ and\
  \citenamefont {R\"opke}}]{PhysRevLett.87.192501}%
  \BibitemOpen
  \bibfield  {author} {\bibinfo {author} {\bibfnamefont {A.}~\bibnamefont
  {Tohsaki}}, \bibinfo {author} {\bibfnamefont {H.}~\bibnamefont {Horiuchi}},
  \bibinfo {author} {\bibfnamefont {P.}~\bibnamefont {Schuck}}, \ and\ \bibinfo
  {author} {\bibfnamefont {G.}~\bibnamefont {R\"opke}},\ }\href {\doibase
  10.1103/PhysRevLett.87.192501} {\bibfield  {journal} {\bibinfo  {journal}
  {Phys. Rev. Lett.}\ }\textbf {\bibinfo {volume} {87}},\ \bibinfo {pages}
  {192501} (\bibinfo {year} {2001})}\BibitemShut {NoStop}%
\bibitem [{\citenamefont {Binder}\ \emph {et~al.}(2014)\citenamefont {Binder},
  \citenamefont {Langhammer}, \citenamefont {Calci},\ and\ \citenamefont
  {Roth}}]{BINDER2014119}%
  \BibitemOpen
  \bibfield  {author} {\bibinfo {author} {\bibfnamefont {S.}~\bibnamefont
  {Binder}}, \bibinfo {author} {\bibfnamefont {J.}~\bibnamefont {Langhammer}},
  \bibinfo {author} {\bibfnamefont {A.}~\bibnamefont {Calci}}, \ and\ \bibinfo
  {author} {\bibfnamefont {R.}~\bibnamefont {Roth}},\ }\href@noop {} {\bibfield
   {journal} {\bibinfo  {journal} {Physics Letters B}\ }\textbf {\bibinfo
  {volume} {736}},\ \bibinfo {pages} {119 } (\bibinfo {year}
  {2014})}\BibitemShut {NoStop}%
\bibitem [{\citenamefont {Maris}\ \emph {et~al.}(2009)\citenamefont {Maris},
  \citenamefont {Vary},\ and\ \citenamefont {Shirokov}}]{PhysRevC.79.014308}%
  \BibitemOpen
  \bibfield  {author} {\bibinfo {author} {\bibfnamefont {P.}~\bibnamefont
  {Maris}}, \bibinfo {author} {\bibfnamefont {J.~P.}\ \bibnamefont {Vary}}, \
  and\ \bibinfo {author} {\bibfnamefont {A.~M.}\ \bibnamefont {Shirokov}},\
  }\href {\doibase 10.1103/PhysRevC.79.014308} {\bibfield  {journal} {\bibinfo
  {journal} {Phys. Rev. C}\ }\textbf {\bibinfo {volume} {79}},\ \bibinfo
  {pages} {014308} (\bibinfo {year} {2009})}\BibitemShut {NoStop}%
\bibitem [{\citenamefont {Dreyfuss}\ \emph {et~al.}(2013)\citenamefont
  {Dreyfuss}, \citenamefont {Launey}, \citenamefont {Dytrych}, \citenamefont
  {Draayer},\ and\ \citenamefont {Bahri}}]{DREYFUSS2013511}%
  \BibitemOpen
  \bibfield  {author} {\bibinfo {author} {\bibfnamefont {A.~C.}\ \bibnamefont
  {Dreyfuss}}, \bibinfo {author} {\bibfnamefont {K.~D.}\ \bibnamefont
  {Launey}}, \bibinfo {author} {\bibfnamefont {T.}~\bibnamefont {Dytrych}},
  \bibinfo {author} {\bibfnamefont {J.~P.}\ \bibnamefont {Draayer}}, \ and\
  \bibinfo {author} {\bibfnamefont {C.}~\bibnamefont {Bahri}},\ }\href
  {\doibase https://doi.org/10.1016/j.physletb.2013.10.048} {\bibfield
  {journal} {\bibinfo  {journal} {Physics Letters B}\ }\textbf {\bibinfo
  {volume} {727}},\ \bibinfo {pages} {511 } (\bibinfo {year}
  {2013})}\BibitemShut {NoStop}%
\bibitem [{\citenamefont {Yoshida}\ \emph {et~al.}(2014)\citenamefont
  {Yoshida}, \citenamefont {Shimizu}, \citenamefont {Abe},\ and\ \citenamefont
  {Otsuka}}]{Yoshida_2014}%
  \BibitemOpen
  \bibfield  {author} {\bibinfo {author} {\bibfnamefont {T.}~\bibnamefont
  {Yoshida}}, \bibinfo {author} {\bibfnamefont {N.}~\bibnamefont {Shimizu}},
  \bibinfo {author} {\bibfnamefont {T.}~\bibnamefont {Abe}}, \ and\ \bibinfo
  {author} {\bibfnamefont {T.}~\bibnamefont {Otsuka}},\ }\href {\doibase
  10.1088/1742-6596/569/1/012063} {\bibfield  {journal} {\bibinfo  {journal}
  {Journal of Physics: Conference Series}\ }\textbf {\bibinfo {volume} {569}},\
  \bibinfo {pages} {012063} (\bibinfo {year} {2014})}\BibitemShut {NoStop}%
\bibitem [{\citenamefont {Neff}\ and\ \citenamefont
  {Feldmeier}(2004)}]{NEFF2004357}%
  \BibitemOpen
  \bibfield  {author} {\bibinfo {author} {\bibfnamefont {T.}~\bibnamefont
  {Neff}}\ and\ \bibinfo {author} {\bibfnamefont {H.}~\bibnamefont
  {Feldmeier}},\ }\href {\doibase
  https://doi.org/10.1016/j.nuclphysa.2004.04.061} {\bibfield  {journal}
  {\bibinfo  {journal} {Nuclear Physics A}\ }\textbf {\bibinfo {volume}
  {738}},\ \bibinfo {pages} {357 } (\bibinfo {year} {2004})},\ \bibinfo {note}
  {proceedings of the 8th International Conference on Clustering Aspects of
  Nuclear Structure and Dynamics}\BibitemShut {NoStop}%
\bibitem [{\citenamefont {Chernykh}\ \emph {et~al.}(2010)\citenamefont
  {Chernykh}, \citenamefont {Feldmeier}, \citenamefont {Neff}, \citenamefont
  {von Neumann-Cosel},\ and\ \citenamefont {Richter}}]{PhysRevLett.105.022501}%
  \BibitemOpen
  \bibfield  {author} {\bibinfo {author} {\bibfnamefont {M.}~\bibnamefont
  {Chernykh}}, \bibinfo {author} {\bibfnamefont {H.}~\bibnamefont {Feldmeier}},
  \bibinfo {author} {\bibfnamefont {T.}~\bibnamefont {Neff}}, \bibinfo {author}
  {\bibfnamefont {P.}~\bibnamefont {von Neumann-Cosel}}, \ and\ \bibinfo
  {author} {\bibfnamefont {A.}~\bibnamefont {Richter}},\ }\href {\doibase
  10.1103/PhysRevLett.105.022501} {\bibfield  {journal} {\bibinfo  {journal}
  {Phys. Rev. Lett.}\ }\textbf {\bibinfo {volume} {105}},\ \bibinfo {pages}
  {022501} (\bibinfo {year} {2010})}\BibitemShut {NoStop}%
\bibitem [{\citenamefont {Itagaki}\ \emph {et~al.}(2011)\citenamefont
  {Itagaki}, \citenamefont {Cseh},\ and\ \citenamefont
  {P\l{}oszajczak}}]{PhysRevC.83.014302}%
  \BibitemOpen
  \bibfield  {author} {\bibinfo {author} {\bibfnamefont {N.}~\bibnamefont
  {Itagaki}}, \bibinfo {author} {\bibfnamefont {J.}~\bibnamefont {Cseh}}, \
  and\ \bibinfo {author} {\bibfnamefont {M.}~\bibnamefont {P\l{}oszajczak}},\
  }\href {\doibase 10.1103/PhysRevC.83.014302} {\bibfield  {journal} {\bibinfo
  {journal} {Phys. Rev. C}\ }\textbf {\bibinfo {volume} {83}},\ \bibinfo
  {pages} {014302} (\bibinfo {year} {2011})}\BibitemShut {NoStop}%
\bibitem [{\citenamefont {Itagaki}\ \emph {et~al.}(2016)\citenamefont
  {Itagaki}, \citenamefont {Matsuno},\ and\ \citenamefont
  {Suhara}}]{ptep093D01}%
  \BibitemOpen
  \bibfield  {author} {\bibinfo {author} {\bibfnamefont {N.}~\bibnamefont
  {Itagaki}}, \bibinfo {author} {\bibfnamefont {H.}~\bibnamefont {Matsuno}}, \
  and\ \bibinfo {author} {\bibfnamefont {T.}~\bibnamefont {Suhara}},\
  }\href@noop {} {\bibfield  {journal} {\bibinfo  {journal} {Progress of
  Theoretical and Experimental Physics}\ }\textbf {\bibinfo {volume} {2016}},\
  \bibinfo {pages} {093D01} (\bibinfo {year} {2016})}\BibitemShut {NoStop}%
\bibitem [{\citenamefont {Suhara}\ \emph {et~al.}(2013)\citenamefont {Suhara},
  \citenamefont {Itagaki}, \citenamefont {Cseh},\ and\ \citenamefont
  {P\l{}oszajczak}}]{PhysRevC.87.054334}%
  \BibitemOpen
  \bibfield  {author} {\bibinfo {author} {\bibfnamefont {T.}~\bibnamefont
  {Suhara}}, \bibinfo {author} {\bibfnamefont {N.}~\bibnamefont {Itagaki}},
  \bibinfo {author} {\bibfnamefont {J.}~\bibnamefont {Cseh}}, \ and\ \bibinfo
  {author} {\bibfnamefont {M.}~\bibnamefont {P\l{}oszajczak}},\ }\href
  {\doibase 10.1103/PhysRevC.87.054334} {\bibfield  {journal} {\bibinfo
  {journal} {Phys. Rev. C}\ }\textbf {\bibinfo {volume} {87}},\ \bibinfo
  {pages} {054334} (\bibinfo {year} {2013})}\BibitemShut {NoStop}%
\bibitem [{\citenamefont {Itagaki}(2016)}]{PhysRevC.94.064324}%
  \BibitemOpen
  \bibfield  {author} {\bibinfo {author} {\bibfnamefont {N.}~\bibnamefont
  {Itagaki}},\ }\href {\doibase 10.1103/PhysRevC.94.064324} {\bibfield
  {journal} {\bibinfo  {journal} {Phys. Rev. C}\ }\textbf {\bibinfo {volume}
  {94}},\ \bibinfo {pages} {064324} (\bibinfo {year} {2016})}\BibitemShut
  {NoStop}%
\bibitem [{\citenamefont {Itagaki}\ and\ \citenamefont
  {Tohsaki}(2018{\natexlab{a}})}]{PhysRevC.97.014307}%
  \BibitemOpen
  \bibfield  {author} {\bibinfo {author} {\bibfnamefont {N.}~\bibnamefont
  {Itagaki}}\ and\ \bibinfo {author} {\bibfnamefont {A.}~\bibnamefont
  {Tohsaki}},\ }\href {\doibase 10.1103/PhysRevC.97.014307} {\bibfield
  {journal} {\bibinfo  {journal} {Phys. Rev. C}\ }\textbf {\bibinfo {volume}
  {97}},\ \bibinfo {pages} {014307} (\bibinfo {year}
  {2018}{\natexlab{a}})}\BibitemShut {NoStop}%
\bibitem [{\citenamefont {Itagaki}\ \emph {et~al.}(2018)\citenamefont
  {Itagaki}, \citenamefont {Matsuno},\ and\ \citenamefont
  {Tohsaki}}]{PhysRevC.98.044306}%
  \BibitemOpen
  \bibfield  {author} {\bibinfo {author} {\bibfnamefont {N.}~\bibnamefont
  {Itagaki}}, \bibinfo {author} {\bibfnamefont {H.}~\bibnamefont {Matsuno}}, \
  and\ \bibinfo {author} {\bibfnamefont {A.}~\bibnamefont {Tohsaki}},\ }\href
  {\doibase 10.1103/PhysRevC.98.044306} {\bibfield  {journal} {\bibinfo
  {journal} {Phys. Rev. C}\ }\textbf {\bibinfo {volume} {98}},\ \bibinfo
  {pages} {044306} (\bibinfo {year} {2018})}\BibitemShut {NoStop}%
\bibitem [{\citenamefont {Itagaki}\ \emph {et~al.}(2005)\citenamefont
  {Itagaki}, \citenamefont {Masui}, \citenamefont {Ito},\ and\ \citenamefont
  {Aoyama}}]{PhysRevC.71.064307}%
  \BibitemOpen
  \bibfield  {author} {\bibinfo {author} {\bibfnamefont {N.}~\bibnamefont
  {Itagaki}}, \bibinfo {author} {\bibfnamefont {H.}~\bibnamefont {Masui}},
  \bibinfo {author} {\bibfnamefont {M.}~\bibnamefont {Ito}}, \ and\ \bibinfo
  {author} {\bibfnamefont {S.}~\bibnamefont {Aoyama}},\ }\href {\doibase
  10.1103/PhysRevC.71.064307} {\bibfield  {journal} {\bibinfo  {journal} {Phys.
  Rev. C}\ }\textbf {\bibinfo {volume} {71}},\ \bibinfo {pages} {064307}
  (\bibinfo {year} {2005})}\BibitemShut {NoStop}%
\bibitem [{\citenamefont {Masui}\ and\ \citenamefont
  {Itagaki}(2007)}]{PhysRevC.75.054309}%
  \BibitemOpen
  \bibfield  {author} {\bibinfo {author} {\bibfnamefont {H.}~\bibnamefont
  {Masui}}\ and\ \bibinfo {author} {\bibfnamefont {N.}~\bibnamefont
  {Itagaki}},\ }\href {\doibase 10.1103/PhysRevC.75.054309} {\bibfield
  {journal} {\bibinfo  {journal} {Phys. Rev. C}\ }\textbf {\bibinfo {volume}
  {75}},\ \bibinfo {pages} {054309} (\bibinfo {year} {2007})}\BibitemShut
  {NoStop}%
\bibitem [{\citenamefont {Itagaki}\ \emph {et~al.}(2019)\citenamefont
  {Itagaki}, \citenamefont {Matsuno},\ and\ \citenamefont
  {Kanada-En’yo}}]{PTEP-Itagaki-ptz046}%
  \BibitemOpen
  \bibfield  {author} {\bibinfo {author} {\bibfnamefont {N.}~\bibnamefont
  {Itagaki}}, \bibinfo {author} {\bibfnamefont {H.}~\bibnamefont {Matsuno}}, \
  and\ \bibinfo {author} {\bibfnamefont {Y.}~\bibnamefont {Kanada-En’yo}},\
  }\href {\doibase 10.1093/ptep/ptz046} {\bibfield  {journal} {\bibinfo
  {journal} {Prog. Theor. Exp. Phys.}\ }\textbf {\bibinfo {volume} {2019}}
  (\bibinfo {year} {2019}),\ 10.1093/ptep/ptz046},\ \bibinfo {note}
  {063D02}\BibitemShut {NoStop}%
\bibitem [{\citenamefont {Matsuno}\ \emph {et~al.}(2018)\citenamefont
  {Matsuno}, \citenamefont {Kanada-En'yo},\ and\ \citenamefont
  {Itagaki}}]{PhysRevC.98.054306}%
  \BibitemOpen
  \bibfield  {author} {\bibinfo {author} {\bibfnamefont {H.}~\bibnamefont
  {Matsuno}}, \bibinfo {author} {\bibfnamefont {Y.}~\bibnamefont
  {Kanada-En'yo}}, \ and\ \bibinfo {author} {\bibfnamefont {N.}~\bibnamefont
  {Itagaki}},\ }\href {\doibase 10.1103/PhysRevC.98.054306} {\bibfield
  {journal} {\bibinfo  {journal} {Phys. Rev. C}\ }\textbf {\bibinfo {volume}
  {98}},\ \bibinfo {pages} {054306} (\bibinfo {year} {2018})}\BibitemShut
  {NoStop}%
\bibitem [{\citenamefont {Itagaki}\ \emph
  {et~al.}(2006{\natexlab{a}})\citenamefont {Itagaki}, \citenamefont {Masui},
  \citenamefont {Ito}, \citenamefont {Aoyama},\ and\ \citenamefont
  {Ikeda}}]{PhysRevC.73.034310}%
  \BibitemOpen
  \bibfield  {author} {\bibinfo {author} {\bibfnamefont {N.}~\bibnamefont
  {Itagaki}}, \bibinfo {author} {\bibfnamefont {H.}~\bibnamefont {Masui}},
  \bibinfo {author} {\bibfnamefont {M.}~\bibnamefont {Ito}}, \bibinfo {author}
  {\bibfnamefont {S.}~\bibnamefont {Aoyama}}, \ and\ \bibinfo {author}
  {\bibfnamefont {K.}~\bibnamefont {Ikeda}},\ }\href {\doibase
  10.1103/PhysRevC.73.034310} {\bibfield  {journal} {\bibinfo  {journal} {Phys.
  Rev. C}\ }\textbf {\bibinfo {volume} {73}},\ \bibinfo {pages} {034310}
  (\bibinfo {year} {2006}{\natexlab{a}})}\BibitemShut {NoStop}%
\bibitem [{\citenamefont {Itagaki}\ and\ \citenamefont
  {Tohsaki}(2018{\natexlab{b}})}]{PhysRevC.97.014304}%
  \BibitemOpen
  \bibfield  {author} {\bibinfo {author} {\bibfnamefont {N.}~\bibnamefont
  {Itagaki}}\ and\ \bibinfo {author} {\bibfnamefont {A.}~\bibnamefont
  {Tohsaki}},\ }\href {\doibase 10.1103/PhysRevC.97.014304} {\bibfield
  {journal} {\bibinfo  {journal} {Phys. Rev. C}\ }\textbf {\bibinfo {volume}
  {97}},\ \bibinfo {pages} {014304} (\bibinfo {year}
  {2018}{\natexlab{b}})}\BibitemShut {NoStop}%
\bibitem [{\citenamefont {Tamagaki}(1968)}]{PTP.39.91}%
  \BibitemOpen
  \bibfield  {author} {\bibinfo {author} {\bibfnamefont {R.}~\bibnamefont
  {Tamagaki}},\ }\href@noop {} {\bibfield  {journal} {\bibinfo  {journal}
  {Progress of Theoretical Physics}\ }\textbf {\bibinfo {volume} {39}},\
  \bibinfo {pages} {91} (\bibinfo {year} {1968})}\BibitemShut {NoStop}%
\bibitem [{\citenamefont {Volkov}(1965)}]{VOLKOV196533}%
  \BibitemOpen
  \bibfield  {author} {\bibinfo {author} {\bibfnamefont {A.}~\bibnamefont
  {Volkov}},\ }\href {\doibase https://doi.org/10.1016/0029-5582(65)90244-0}
  {\bibfield  {journal} {\bibinfo  {journal} {Nuclear Physics}\ }\textbf
  {\bibinfo {volume} {74}},\ \bibinfo {pages} {33 } (\bibinfo {year}
  {1965})}\BibitemShut {NoStop}%
\bibitem [{\citenamefont {Tohsaki}(1994)}]{PhysRevC.49.1814}%
  \BibitemOpen
  \bibfield  {author} {\bibinfo {author} {\bibfnamefont {A.}~\bibnamefont
  {Tohsaki}},\ }\href {\doibase 10.1103/PhysRevC.49.1814} {\bibfield  {journal}
  {\bibinfo  {journal} {Phys. Rev. C}\ }\textbf {\bibinfo {volume} {49}},\
  \bibinfo {pages} {1814} (\bibinfo {year} {1994})}\BibitemShut {NoStop}%
\bibitem [{\citenamefont {Itagaki}\ \emph {et~al.}(1995)\citenamefont
  {Itagaki}, \citenamefont {Ohnishi},\ and\ \citenamefont
  {Kato}}]{PTP.94.1019}%
  \BibitemOpen
  \bibfield  {author} {\bibinfo {author} {\bibfnamefont {N.}~\bibnamefont
  {Itagaki}}, \bibinfo {author} {\bibfnamefont {A.}~\bibnamefont {Ohnishi}}, \
  and\ \bibinfo {author} {\bibfnamefont {K.}~\bibnamefont {Kato}},\ }\href
  {\doibase 10.1143/PTP.94.1019} {\bibfield  {journal} {\bibinfo  {journal}
  {Progress of Theoretical Physics}\ }\textbf {\bibinfo {volume} {94}},\
  \bibinfo {pages} {1019} (\bibinfo {year} {1995})}\BibitemShut {NoStop}%
\bibitem [{\citenamefont {Freer}\ \emph {et~al.}(2018)\citenamefont {Freer},
  \citenamefont {Horiuchi}, \citenamefont {Kanada-En'yo}, \citenamefont {Lee},\
  and\ \citenamefont {Mei\ss{}ner}}]{RevModPhys.90.035004}%
  \BibitemOpen
  \bibfield  {author} {\bibinfo {author} {\bibfnamefont {M.}~\bibnamefont
  {Freer}}, \bibinfo {author} {\bibfnamefont {H.}~\bibnamefont {Horiuchi}},
  \bibinfo {author} {\bibfnamefont {Y.}~\bibnamefont {Kanada-En'yo}}, \bibinfo
  {author} {\bibfnamefont {D.}~\bibnamefont {Lee}}, \ and\ \bibinfo {author}
  {\bibfnamefont {U.-G.}\ \bibnamefont {Mei\ss{}ner}},\ }\href {\doibase
  10.1103/RevModPhys.90.035004} {\bibfield  {journal} {\bibinfo  {journal}
  {Rev. Mod. Phys.}\ }\textbf {\bibinfo {volume} {90}},\ \bibinfo {pages}
  {035004} (\bibinfo {year} {2018})}\BibitemShut {NoStop}%
\bibitem [{\citenamefont {Morinaga}(1956)}]{PhysRev.101.254}%
  \BibitemOpen
  \bibfield  {author} {\bibinfo {author} {\bibfnamefont {H.}~\bibnamefont
  {Morinaga}},\ }\href {\doibase 10.1103/PhysRev.101.254} {\bibfield  {journal}
  {\bibinfo  {journal} {Phys. Rev.}\ }\textbf {\bibinfo {volume} {101}},\
  \bibinfo {pages} {254} (\bibinfo {year} {1956})}\BibitemShut {NoStop}%
\bibitem [{\citenamefont {Itoh}\ \emph {et~al.}(2011)\citenamefont {Itoh},
  \citenamefont {Akimune}, \citenamefont {Fujiwara}, \citenamefont {Garg},
  \citenamefont {Hashimoto}, \citenamefont {Kawabata}, \citenamefont {Kawase},
  \citenamefont {Kishi}, \citenamefont {Murakami}, \citenamefont {Nakanishi},
  \citenamefont {Nakatsugawa}, \citenamefont {Nayak}, \citenamefont {Okumura},
  \citenamefont {Sakaguchi}, \citenamefont {Takeda}, \citenamefont {Terashima},
  \citenamefont {Uchida}, \citenamefont {Yasuda}, \citenamefont {Yosoi},\ and\
  \citenamefont {Zenihiro}}]{PhysRevC.84.054308}%
  \BibitemOpen
  \bibfield  {author} {\bibinfo {author} {\bibfnamefont {M.}~\bibnamefont
  {Itoh}}, \bibinfo {author} {\bibfnamefont {H.}~\bibnamefont {Akimune}},
  \bibinfo {author} {\bibfnamefont {M.}~\bibnamefont {Fujiwara}}, \bibinfo
  {author} {\bibfnamefont {U.}~\bibnamefont {Garg}}, \bibinfo {author}
  {\bibfnamefont {N.}~\bibnamefont {Hashimoto}}, \bibinfo {author}
  {\bibfnamefont {T.}~\bibnamefont {Kawabata}}, \bibinfo {author}
  {\bibfnamefont {K.}~\bibnamefont {Kawase}}, \bibinfo {author} {\bibfnamefont
  {S.}~\bibnamefont {Kishi}}, \bibinfo {author} {\bibfnamefont
  {T.}~\bibnamefont {Murakami}}, \bibinfo {author} {\bibfnamefont
  {K.}~\bibnamefont {Nakanishi}}, \bibinfo {author} {\bibfnamefont
  {Y.}~\bibnamefont {Nakatsugawa}}, \bibinfo {author} {\bibfnamefont {B.~K.}\
  \bibnamefont {Nayak}}, \bibinfo {author} {\bibfnamefont {S.}~\bibnamefont
  {Okumura}}, \bibinfo {author} {\bibfnamefont {H.}~\bibnamefont {Sakaguchi}},
  \bibinfo {author} {\bibfnamefont {H.}~\bibnamefont {Takeda}}, \bibinfo
  {author} {\bibfnamefont {S.}~\bibnamefont {Terashima}}, \bibinfo {author}
  {\bibfnamefont {M.}~\bibnamefont {Uchida}}, \bibinfo {author} {\bibfnamefont
  {Y.}~\bibnamefont {Yasuda}}, \bibinfo {author} {\bibfnamefont
  {M.}~\bibnamefont {Yosoi}}, \ and\ \bibinfo {author} {\bibfnamefont
  {J.}~\bibnamefont {Zenihiro}},\ }\href {\doibase 10.1103/PhysRevC.84.054308}
  {\bibfield  {journal} {\bibinfo  {journal} {Phys. Rev. C}\ }\textbf {\bibinfo
  {volume} {84}},\ \bibinfo {pages} {054308} (\bibinfo {year}
  {2011})}\BibitemShut {NoStop}%
\bibitem [{\citenamefont {Suhara}\ and\ \citenamefont
  {Kanada-En'yo}(2015)}]{PhysRevC.91.024315}%
  \BibitemOpen
  \bibfield  {author} {\bibinfo {author} {\bibfnamefont {T.}~\bibnamefont
  {Suhara}}\ and\ \bibinfo {author} {\bibfnamefont {Y.}~\bibnamefont
  {Kanada-En'yo}},\ }\href {\doibase 10.1103/PhysRevC.91.024315} {\bibfield
  {journal} {\bibinfo  {journal} {Phys. Rev. C}\ }\textbf {\bibinfo {volume}
  {91}},\ \bibinfo {pages} {024315} (\bibinfo {year} {2015})}\BibitemShut
  {NoStop}%
\bibitem [{\citenamefont {Chevallier}\ \emph {et~al.}(1967)\citenamefont
  {Chevallier}, \citenamefont {Scheibling}, \citenamefont {Goldring},
  \citenamefont {Plesser},\ and\ \citenamefont {Sachs}}]{PhysRev.160.827}%
  \BibitemOpen
  \bibfield  {author} {\bibinfo {author} {\bibfnamefont {P.}~\bibnamefont
  {Chevallier}}, \bibinfo {author} {\bibfnamefont {F.}~\bibnamefont
  {Scheibling}}, \bibinfo {author} {\bibfnamefont {G.}~\bibnamefont
  {Goldring}}, \bibinfo {author} {\bibfnamefont {I.}~\bibnamefont {Plesser}}, \
  and\ \bibinfo {author} {\bibfnamefont {M.~W.}\ \bibnamefont {Sachs}},\ }\href
  {\doibase 10.1103/PhysRev.160.827} {\bibfield  {journal} {\bibinfo  {journal}
  {Phys. Rev.}\ }\textbf {\bibinfo {volume} {160}},\ \bibinfo {pages} {827}
  (\bibinfo {year} {1967})}\BibitemShut {NoStop}%
\bibitem [{\citenamefont {Wuosmaa}\ \emph {et~al.}(1992)\citenamefont
  {Wuosmaa}, \citenamefont {Betts}, \citenamefont {Back}, \citenamefont
  {Freer}, \citenamefont {Glagola}, \citenamefont {Happ}, \citenamefont
  {Henderson}, \citenamefont {Wilt},\ and\ \citenamefont
  {Bearden}}]{PhysRevLett.68.1295}%
  \BibitemOpen
  \bibfield  {author} {\bibinfo {author} {\bibfnamefont {A.~H.}\ \bibnamefont
  {Wuosmaa}}, \bibinfo {author} {\bibfnamefont {R.~R.}\ \bibnamefont {Betts}},
  \bibinfo {author} {\bibfnamefont {B.~B.}\ \bibnamefont {Back}}, \bibinfo
  {author} {\bibfnamefont {M.}~\bibnamefont {Freer}}, \bibinfo {author}
  {\bibfnamefont {B.~G.}\ \bibnamefont {Glagola}}, \bibinfo {author}
  {\bibfnamefont {T.}~\bibnamefont {Happ}}, \bibinfo {author} {\bibfnamefont
  {D.~J.}\ \bibnamefont {Henderson}}, \bibinfo {author} {\bibfnamefont
  {P.}~\bibnamefont {Wilt}}, \ and\ \bibinfo {author} {\bibfnamefont {I.~G.}\
  \bibnamefont {Bearden}},\ }\href {\doibase 10.1103/PhysRevLett.68.1295}
  {\bibfield  {journal} {\bibinfo  {journal} {Phys. Rev. Lett.}\ }\textbf
  {\bibinfo {volume} {68}},\ \bibinfo {pages} {1295} (\bibinfo {year}
  {1992})}\BibitemShut {NoStop}%
\bibitem [{\citenamefont {Flocard}\ \emph {et~al.}(1984)\citenamefont
  {Flocard}, \citenamefont {Heenen}, \citenamefont {Krieger},\ and\
  \citenamefont {Weiss}}]{Flocard-PTP.72.1000}%
  \BibitemOpen
  \bibfield  {author} {\bibinfo {author} {\bibfnamefont {H.}~\bibnamefont
  {Flocard}}, \bibinfo {author} {\bibfnamefont {P.~H.}\ \bibnamefont {Heenen}},
  \bibinfo {author} {\bibfnamefont {S.~J.}\ \bibnamefont {Krieger}}, \ and\
  \bibinfo {author} {\bibfnamefont {M.~S.}\ \bibnamefont {Weiss}},\ }\href
  {\doibase 10.1143/PTP.72.1000} {\bibfield  {journal} {\bibinfo  {journal}
  {Prog. Theor. Phys.}\ }\textbf {\bibinfo {volume} {72}},\ \bibinfo {pages}
  {1000} (\bibinfo {year} {1984})}\BibitemShut {NoStop}%
\bibitem [{\citenamefont {Itagaki}\ \emph {et~al.}(2001)\citenamefont
  {Itagaki}, \citenamefont {Okabe}, \citenamefont {Ikeda},\ and\ \citenamefont
  {Tanihata}}]{PhysRevC.64.014301}%
  \BibitemOpen
  \bibfield  {author} {\bibinfo {author} {\bibfnamefont {N.}~\bibnamefont
  {Itagaki}}, \bibinfo {author} {\bibfnamefont {S.}~\bibnamefont {Okabe}},
  \bibinfo {author} {\bibfnamefont {K.}~\bibnamefont {Ikeda}}, \ and\ \bibinfo
  {author} {\bibfnamefont {I.}~\bibnamefont {Tanihata}},\ }\href {\doibase
  10.1103/PhysRevC.64.014301} {\bibfield  {journal} {\bibinfo  {journal} {Phys.
  Rev. C}\ }\textbf {\bibinfo {volume} {64}},\ \bibinfo {pages} {014301}
  (\bibinfo {year} {2001})}\BibitemShut {NoStop}%
\bibitem [{\citenamefont {Bender}\ and\ \citenamefont
  {Heenen}(2003)}]{BENDER2003390}%
  \BibitemOpen
  \bibfield  {author} {\bibinfo {author} {\bibfnamefont {M.}~\bibnamefont
  {Bender}}\ and\ \bibinfo {author} {\bibfnamefont {P.-H.}\ \bibnamefont
  {Heenen}},\ }\href {\doibase https://doi.org/10.1016/S0375-9474(02)01308-8}
  {\bibfield  {journal} {\bibinfo  {journal} {Nuclear Physics A}\ }\textbf
  {\bibinfo {volume} {713}},\ \bibinfo {pages} {390 } (\bibinfo {year}
  {2003})}\BibitemShut {NoStop}%
\bibitem [{\citenamefont {Itagaki}\ \emph
  {et~al.}(2006{\natexlab{b}})\citenamefont {Itagaki}, \citenamefont {von
  Oertzen},\ and\ \citenamefont {Okabe}}]{PhysRevC.74.067304}%
  \BibitemOpen
  \bibfield  {author} {\bibinfo {author} {\bibfnamefont {N.}~\bibnamefont
  {Itagaki}}, \bibinfo {author} {\bibfnamefont {W.}~\bibnamefont {von
  Oertzen}}, \ and\ \bibinfo {author} {\bibfnamefont {S.}~\bibnamefont
  {Okabe}},\ }\href {\doibase 10.1103/PhysRevC.74.067304} {\bibfield  {journal}
  {\bibinfo  {journal} {Phys. Rev. C}\ }\textbf {\bibinfo {volume} {74}},\
  \bibinfo {pages} {067304} (\bibinfo {year} {2006}{\natexlab{b}})}\BibitemShut
  {NoStop}%
\bibitem [{\citenamefont {Maruhn}\ \emph {et~al.}(2010)\citenamefont {Maruhn},
  \citenamefont {Loebl}, \citenamefont {Itagaki},\ and\ \citenamefont
  {Kimura}}]{MARUHN20101}%
  \BibitemOpen
  \bibfield  {author} {\bibinfo {author} {\bibfnamefont {J.}~\bibnamefont
  {Maruhn}}, \bibinfo {author} {\bibfnamefont {N.}~\bibnamefont {Loebl}},
  \bibinfo {author} {\bibfnamefont {N.}~\bibnamefont {Itagaki}}, \ and\
  \bibinfo {author} {\bibfnamefont {M.}~\bibnamefont {Kimura}},\ }\href
  {\doibase https://doi.org/10.1016/j.nuclphysa.2009.12.005} {\bibfield
  {journal} {\bibinfo  {journal} {Nuclear Physics A}\ }\textbf {\bibinfo
  {volume} {833}},\ \bibinfo {pages} {1 } (\bibinfo {year} {2010})}\BibitemShut
  {NoStop}%
\bibitem [{\citenamefont {Suhara}\ and\ \citenamefont
  {Kanada-En'yo}(2010)}]{PhysRevC.82.044301}%
  \BibitemOpen
  \bibfield  {author} {\bibinfo {author} {\bibfnamefont {T.}~\bibnamefont
  {Suhara}}\ and\ \bibinfo {author} {\bibfnamefont {Y.}~\bibnamefont
  {Kanada-En'yo}},\ }\href {\doibase 10.1103/PhysRevC.82.044301} {\bibfield
  {journal} {\bibinfo  {journal} {Phys. Rev. C}\ }\textbf {\bibinfo {volume}
  {82}},\ \bibinfo {pages} {044301} (\bibinfo {year} {2010})}\BibitemShut
  {NoStop}%
\bibitem [{\citenamefont {Furutachi}\ and\ \citenamefont
  {Kimura}(2011)}]{PhysRevC.83.021303}%
  \BibitemOpen
  \bibfield  {author} {\bibinfo {author} {\bibfnamefont {N.}~\bibnamefont
  {Furutachi}}\ and\ \bibinfo {author} {\bibfnamefont {M.}~\bibnamefont
  {Kimura}},\ }\href {\doibase 10.1103/PhysRevC.83.021303} {\bibfield
  {journal} {\bibinfo  {journal} {Phys. Rev. C}\ }\textbf {\bibinfo {volume}
  {83}},\ \bibinfo {pages} {021303} (\bibinfo {year} {2011})}\BibitemShut
  {NoStop}%
\bibitem [{\citenamefont {Ichikawa}\ \emph {et~al.}(2011)\citenamefont
  {Ichikawa}, \citenamefont {Maruhn}, \citenamefont {Itagaki},\ and\
  \citenamefont {Ohkubo}}]{PhysRevLett.107.112501}%
  \BibitemOpen
  \bibfield  {author} {\bibinfo {author} {\bibfnamefont {T.}~\bibnamefont
  {Ichikawa}}, \bibinfo {author} {\bibfnamefont {J.~A.}\ \bibnamefont
  {Maruhn}}, \bibinfo {author} {\bibfnamefont {N.}~\bibnamefont {Itagaki}}, \
  and\ \bibinfo {author} {\bibfnamefont {S.}~\bibnamefont {Ohkubo}},\ }\href
  {\doibase 10.1103/PhysRevLett.107.112501} {\bibfield  {journal} {\bibinfo
  {journal} {Phys. Rev. Lett.}\ }\textbf {\bibinfo {volume} {107}},\ \bibinfo
  {pages} {112501} (\bibinfo {year} {2011})}\BibitemShut {NoStop}%
\bibitem [{\citenamefont {Yao}\ \emph {et~al.}(2014)\citenamefont {Yao},
  \citenamefont {Itagaki},\ and\ \citenamefont {Meng}}]{PhysRevC.90.054307}%
  \BibitemOpen
  \bibfield  {author} {\bibinfo {author} {\bibfnamefont {J.~M.}\ \bibnamefont
  {Yao}}, \bibinfo {author} {\bibfnamefont {N.}~\bibnamefont {Itagaki}}, \ and\
  \bibinfo {author} {\bibfnamefont {J.}~\bibnamefont {Meng}},\ }\href {\doibase
  10.1103/PhysRevC.90.054307} {\bibfield  {journal} {\bibinfo  {journal} {Phys.
  Rev. C}\ }\textbf {\bibinfo {volume} {90}},\ \bibinfo {pages} {054307}
  (\bibinfo {year} {2014})}\BibitemShut {NoStop}%
\bibitem [{\citenamefont {Iwata}\ \emph {et~al.}(2015)\citenamefont {Iwata},
  \citenamefont {Ichikawa}, \citenamefont {Itagaki}, \citenamefont {Maruhn},\
  and\ \citenamefont {Otsuka}}]{PhysRevC.92.011303}%
  \BibitemOpen
  \bibfield  {author} {\bibinfo {author} {\bibfnamefont {Y.}~\bibnamefont
  {Iwata}}, \bibinfo {author} {\bibfnamefont {T.}~\bibnamefont {Ichikawa}},
  \bibinfo {author} {\bibfnamefont {N.}~\bibnamefont {Itagaki}}, \bibinfo
  {author} {\bibfnamefont {J.~A.}\ \bibnamefont {Maruhn}}, \ and\ \bibinfo
  {author} {\bibfnamefont {T.}~\bibnamefont {Otsuka}},\ }\href {\doibase
  10.1103/PhysRevC.92.011303} {\bibfield  {journal} {\bibinfo  {journal} {Phys.
  Rev. C}\ }\textbf {\bibinfo {volume} {92}},\ \bibinfo {pages} {011303}
  (\bibinfo {year} {2015})}\BibitemShut {NoStop}%
\bibitem [{\citenamefont {Suhara}\ \emph {et~al.}(2014)\citenamefont {Suhara},
  \citenamefont {Funaki}, \citenamefont {Zhou}, \citenamefont {Horiuchi},\ and\
  \citenamefont {Tohsaki}}]{PhysRevLett.112.062501}%
  \BibitemOpen
  \bibfield  {author} {\bibinfo {author} {\bibfnamefont {T.}~\bibnamefont
  {Suhara}}, \bibinfo {author} {\bibfnamefont {Y.}~\bibnamefont {Funaki}},
  \bibinfo {author} {\bibfnamefont {B.}~\bibnamefont {Zhou}}, \bibinfo {author}
  {\bibfnamefont {H.}~\bibnamefont {Horiuchi}}, \ and\ \bibinfo {author}
  {\bibfnamefont {A.}~\bibnamefont {Tohsaki}},\ }\href {\doibase
  10.1103/PhysRevLett.112.062501} {\bibfield  {journal} {\bibinfo  {journal}
  {Phys. Rev. Lett.}\ }\textbf {\bibinfo {volume} {112}},\ \bibinfo {pages}
  {062501} (\bibinfo {year} {2014})}\BibitemShut {NoStop}%
\bibitem [{\citenamefont {Zhao}\ \emph {et~al.}(2015)\citenamefont {Zhao},
  \citenamefont {Itagaki},\ and\ \citenamefont
  {Meng}}]{PhysRevLett.115.022501}%
  \BibitemOpen
  \bibfield  {author} {\bibinfo {author} {\bibfnamefont {P.~W.}\ \bibnamefont
  {Zhao}}, \bibinfo {author} {\bibfnamefont {N.}~\bibnamefont {Itagaki}}, \
  and\ \bibinfo {author} {\bibfnamefont {J.}~\bibnamefont {Meng}},\ }\href
  {\doibase 10.1103/PhysRevLett.115.022501} {\bibfield  {journal} {\bibinfo
  {journal} {Phys. Rev. Lett.}\ }\textbf {\bibinfo {volume} {115}},\ \bibinfo
  {pages} {022501} (\bibinfo {year} {2015})}\BibitemShut {NoStop}%
\bibitem [{\citenamefont {Suzuki}\ and\ \citenamefont
  {Horiuchi}(2017)}]{PhysRevC.95.044320}%
  \BibitemOpen
  \bibfield  {author} {\bibinfo {author} {\bibfnamefont {Y.}~\bibnamefont
  {Suzuki}}\ and\ \bibinfo {author} {\bibfnamefont {W.}~\bibnamefont
  {Horiuchi}},\ }\href {\doibase 10.1103/PhysRevC.95.044320} {\bibfield
  {journal} {\bibinfo  {journal} {Phys. Rev. C}\ }\textbf {\bibinfo {volume}
  {95}},\ \bibinfo {pages} {044320} (\bibinfo {year} {2017})}\BibitemShut
  {NoStop}%
\bibitem [{\citenamefont {Baba}\ and\ \citenamefont
  {Kimura}(2018)}]{PhysRevC.97.054315}%
  \BibitemOpen
  \bibfield  {author} {\bibinfo {author} {\bibfnamefont {T.}~\bibnamefont
  {Baba}}\ and\ \bibinfo {author} {\bibfnamefont {M.}~\bibnamefont {Kimura}},\
  }\href {\doibase 10.1103/PhysRevC.97.054315} {\bibfield  {journal} {\bibinfo
  {journal} {Phys. Rev. C}\ }\textbf {\bibinfo {volume} {97}},\ \bibinfo
  {pages} {054315} (\bibinfo {year} {2018})}\BibitemShut {NoStop}%
\bibitem [{\citenamefont {Inakura}\ and\ \citenamefont
  {Mizutori}(2018)}]{PhysRevC.98.044312}%
  \BibitemOpen
  \bibfield  {author} {\bibinfo {author} {\bibfnamefont {T.}~\bibnamefont
  {Inakura}}\ and\ \bibinfo {author} {\bibfnamefont {S.}~\bibnamefont
  {Mizutori}},\ }\href {\doibase 10.1103/PhysRevC.98.044312} {\bibfield
  {journal} {\bibinfo  {journal} {Phys. Rev. C}\ }\textbf {\bibinfo {volume}
  {98}},\ \bibinfo {pages} {044312} (\bibinfo {year} {2018})}\BibitemShut
  {NoStop}%
\bibitem [{\citenamefont {Ren}\ \emph {et~al.}(2019)\citenamefont {Ren},
  \citenamefont {Zhang}, \citenamefont {Zhao}, \citenamefont {Itagaki},
  \citenamefont {Maruhn},\ and\ \citenamefont {Meng}}]{Ren2019}%
  \BibitemOpen
  \bibfield  {author} {\bibinfo {author} {\bibfnamefont {Z.~X.}\ \bibnamefont
  {Ren}}, \bibinfo {author} {\bibfnamefont {S.~Q.}\ \bibnamefont {Zhang}},
  \bibinfo {author} {\bibfnamefont {P.~W.}\ \bibnamefont {Zhao}}, \bibinfo
  {author} {\bibfnamefont {N.}~\bibnamefont {Itagaki}}, \bibinfo {author}
  {\bibfnamefont {J.~A.}\ \bibnamefont {Maruhn}}, \ and\ \bibinfo {author}
  {\bibfnamefont {J.}~\bibnamefont {Meng}},\ }\href {\doibase
  10.1007/s11433-019-9412-3} {\bibfield  {journal} {\bibinfo  {journal} {Sci.
  China Phys, Mechanics {\&} Astronomy}\ }\textbf {\bibinfo {volume} {62}},\
  \bibinfo {pages} {112062} (\bibinfo {year} {2019})}\BibitemShut {NoStop}%
\bibitem [{\citenamefont {Machleidt}\ and\ \citenamefont
  {Entem}(2011)}]{MACHLEIDT20111}%
  \BibitemOpen
  \bibfield  {author} {\bibinfo {author} {\bibfnamefont {R.}~\bibnamefont
  {Machleidt}}\ and\ \bibinfo {author} {\bibfnamefont {D.}~\bibnamefont
  {Entem}},\ }\href {\doibase https://doi.org/10.1016/j.physrep.2011.02.001}
  {\bibfield  {journal} {\bibinfo  {journal} {Physics Reports}\ }\textbf
  {\bibinfo {volume} {503}},\ \bibinfo {pages} {1 } (\bibinfo {year}
  {2011})}\BibitemShut {NoStop}%
\bibitem [{\citenamefont {Mito}\ and\ \citenamefont
  {Kamimura}(1976)}]{Mito.PTP.56.583}%
  \BibitemOpen
  \bibfield  {author} {\bibinfo {author} {\bibfnamefont {Y.}~\bibnamefont
  {Mito}}\ and\ \bibinfo {author} {\bibfnamefont {M.}~\bibnamefont
  {Kamimura}},\ }\href {\doibase 10.1143/PTP.56.583} {\bibfield  {journal}
  {\bibinfo  {journal} {Progress of Theoretical Physics}\ }\textbf {\bibinfo
  {volume} {56}},\ \bibinfo {pages} {583} (\bibinfo {year} {1976})}\BibitemShut
  {NoStop}%
\bibitem [{\citenamefont {Kamimura}(1977)}]{Kamimura.PTPS.62.236}%
  \BibitemOpen
  \bibfield  {author} {\bibinfo {author} {\bibfnamefont {M.}~\bibnamefont
  {Kamimura}},\ }\href {\doibase 10.1143/PTPS.62.236} {\bibfield  {journal}
  {\bibinfo  {journal} {Prog. Theor. Phys. Suppl.}\ }\textbf {\bibinfo {volume}
  {62}},\ \bibinfo {pages} {236} (\bibinfo {year} {1977})}\BibitemShut
  {NoStop}%
\bibitem [{\citenamefont {Tanabe}\ \emph {et~al.}(1975)\citenamefont {Tanabe},
  \citenamefont {Tohsaki},\ and\ \citenamefont
  {Tamagaki}}]{10.1143/PTP.53.677}%
  \BibitemOpen
  \bibfield  {author} {\bibinfo {author} {\bibfnamefont {F.}~\bibnamefont
  {Tanabe}}, \bibinfo {author} {\bibfnamefont {A.}~\bibnamefont {Tohsaki}}, \
  and\ \bibinfo {author} {\bibfnamefont {R.}~\bibnamefont {Tamagaki}},\ }\href
  {\doibase 10.1143/PTP.53.677} {\bibfield  {journal} {\bibinfo  {journal}
  {Progress of Theoretical Physics}\ }\textbf {\bibinfo {volume} {53}},\
  \bibinfo {pages} {677} (\bibinfo {year} {1975})}\BibitemShut {NoStop}%
\bibitem [{\citenamefont {Heydenburg}\ and\ \citenamefont
  {Temmer}(1956)}]{PhysRev.104.123}%
  \BibitemOpen
  \bibfield  {author} {\bibinfo {author} {\bibfnamefont {N.~P.}\ \bibnamefont
  {Heydenburg}}\ and\ \bibinfo {author} {\bibfnamefont {G.~M.}\ \bibnamefont
  {Temmer}},\ }\href {\doibase 10.1103/PhysRev.104.123} {\bibfield  {journal}
  {\bibinfo  {journal} {Phys. Rev.}\ }\textbf {\bibinfo {volume} {104}},\
  \bibinfo {pages} {123} (\bibinfo {year} {1956})}\BibitemShut {NoStop}%
\bibitem [{\citenamefont {Russell}\ \emph {et~al.}(1956)\citenamefont
  {Russell}, \citenamefont {Phillips},\ and\ \citenamefont
  {Reich}}]{PhysRev.104.135}%
  \BibitemOpen
  \bibfield  {author} {\bibinfo {author} {\bibfnamefont {J.~L.}\ \bibnamefont
  {Russell}}, \bibinfo {author} {\bibfnamefont {G.~C.}\ \bibnamefont
  {Phillips}}, \ and\ \bibinfo {author} {\bibfnamefont {C.~W.}\ \bibnamefont
  {Reich}},\ }\href {\doibase 10.1103/PhysRev.104.135} {\bibfield  {journal}
  {\bibinfo  {journal} {Phys. Rev.}\ }\textbf {\bibinfo {volume} {104}},\
  \bibinfo {pages} {135} (\bibinfo {year} {1956})}\BibitemShut {NoStop}%
\bibitem [{\citenamefont {Nilson}\ \emph {et~al.}(1958)\citenamefont {Nilson},
  \citenamefont {Jentschke}, \citenamefont {Briggs}, \citenamefont {Kerman},\
  and\ \citenamefont {Snyder}}]{PhysRev.109.850}%
  \BibitemOpen
  \bibfield  {author} {\bibinfo {author} {\bibfnamefont {R.}~\bibnamefont
  {Nilson}}, \bibinfo {author} {\bibfnamefont {W.~K.}\ \bibnamefont
  {Jentschke}}, \bibinfo {author} {\bibfnamefont {G.~R.}\ \bibnamefont
  {Briggs}}, \bibinfo {author} {\bibfnamefont {R.~O.}\ \bibnamefont {Kerman}},
  \ and\ \bibinfo {author} {\bibfnamefont {J.~N.}\ \bibnamefont {Snyder}},\
  }\href {\doibase 10.1103/PhysRev.109.850} {\bibfield  {journal} {\bibinfo
  {journal} {Phys. Rev.}\ }\textbf {\bibinfo {volume} {109}},\ \bibinfo {pages}
  {850} (\bibinfo {year} {1958})}\BibitemShut {NoStop}%
\bibitem [{\citenamefont {Jones}\ \emph {et~al.}(1960)\citenamefont {Jones},
  \citenamefont {Phillips},\ and\ \citenamefont {Miller}}]{PhysRev.117.525}%
  \BibitemOpen
  \bibfield  {author} {\bibinfo {author} {\bibfnamefont {C.~M.}\ \bibnamefont
  {Jones}}, \bibinfo {author} {\bibfnamefont {G.~C.}\ \bibnamefont {Phillips}},
  \ and\ \bibinfo {author} {\bibfnamefont {P.~D.}\ \bibnamefont {Miller}},\
  }\href {\doibase 10.1103/PhysRev.117.525} {\bibfield  {journal} {\bibinfo
  {journal} {Phys. Rev.}\ }\textbf {\bibinfo {volume} {117}},\ \bibinfo {pages}
  {525} (\bibinfo {year} {1960})}\BibitemShut {NoStop}%
\bibitem [{\citenamefont {Miyake}(1961)}]{MiyakeBICR1961}%
  \BibitemOpen
  \bibfield  {author} {\bibinfo {author} {\bibfnamefont {K.}~\bibnamefont
  {Miyake}},\ }\href@noop {} {\bibfield  {journal} {\bibinfo  {journal}
  {Bulletin of the Institute for Chemical Research, Kyoto University}\ }\textbf
  {\bibinfo {volume} {39}},\ \bibinfo {pages} {313} (\bibinfo {year}
  {1961})}\BibitemShut {NoStop}%
\bibitem [{\citenamefont {Tombrello}\ and\ \citenamefont
  {Senhouse}(1963)}]{PhysRev.129.2252}%
  \BibitemOpen
  \bibfield  {author} {\bibinfo {author} {\bibfnamefont {T.~A.}\ \bibnamefont
  {Tombrello}}\ and\ \bibinfo {author} {\bibfnamefont {L.~S.}\ \bibnamefont
  {Senhouse}},\ }\href {\doibase 10.1103/PhysRev.129.2252} {\bibfield
  {journal} {\bibinfo  {journal} {Phys. Rev.}\ }\textbf {\bibinfo {volume}
  {129}},\ \bibinfo {pages} {2252} (\bibinfo {year} {1963})}\BibitemShut
  {NoStop}%
\bibitem [{\citenamefont {Bacher}\ \emph {et~al.}(1972)\citenamefont {Bacher},
  \citenamefont {Resmini}, \citenamefont {Conzett}, \citenamefont
  {de~Swiniarski}, \citenamefont {Meiner},\ and\ \citenamefont
  {Ernst}}]{PhysRevLett.29.1331}%
  \BibitemOpen
  \bibfield  {author} {\bibinfo {author} {\bibfnamefont {A.~D.}\ \bibnamefont
  {Bacher}}, \bibinfo {author} {\bibfnamefont {F.~G.}\ \bibnamefont {Resmini}},
  \bibinfo {author} {\bibfnamefont {H.~E.}\ \bibnamefont {Conzett}}, \bibinfo
  {author} {\bibfnamefont {R.}~\bibnamefont {de~Swiniarski}}, \bibinfo {author}
  {\bibfnamefont {H.}~\bibnamefont {Meiner}}, \ and\ \bibinfo {author}
  {\bibfnamefont {J.}~\bibnamefont {Ernst}},\ }\href {\doibase
  10.1103/PhysRevLett.29.1331} {\bibfield  {journal} {\bibinfo  {journal}
  {Phys. Rev. Lett.}\ }\textbf {\bibinfo {volume} {29}},\ \bibinfo {pages}
  {1331} (\bibinfo {year} {1972})}\BibitemShut {NoStop}%
\bibitem [{\citenamefont {Freer}\ \emph {et~al.}(2014)\citenamefont {Freer},
  \citenamefont {Malcolm}, \citenamefont {Achouri}, \citenamefont {Ashwood},
  \citenamefont {Bardayan}, \citenamefont {Brown}, \citenamefont {Catford},
  \citenamefont {Chipps}, \citenamefont {Cizewski}, \citenamefont {Curtis},
  \citenamefont {Jones}, \citenamefont {Munoz-Britton}, \citenamefont {Pain},
  \citenamefont {Soi\ifmmode~\acute{c}\else \'{c}\fi{}}, \citenamefont
  {Wheldon}, \citenamefont {Wilson},\ and\ \citenamefont
  {Ziman}}]{PhysRevC.90.054324}%
  \BibitemOpen
  \bibfield  {author} {\bibinfo {author} {\bibfnamefont {M.}~\bibnamefont
  {Freer}}, \bibinfo {author} {\bibfnamefont {J.~D.}\ \bibnamefont {Malcolm}},
  \bibinfo {author} {\bibfnamefont {N.~L.}\ \bibnamefont {Achouri}}, \bibinfo
  {author} {\bibfnamefont {N.~I.}\ \bibnamefont {Ashwood}}, \bibinfo {author}
  {\bibfnamefont {D.~W.}\ \bibnamefont {Bardayan}}, \bibinfo {author}
  {\bibfnamefont {S.~M.}\ \bibnamefont {Brown}}, \bibinfo {author}
  {\bibfnamefont {W.~N.}\ \bibnamefont {Catford}}, \bibinfo {author}
  {\bibfnamefont {K.~A.}\ \bibnamefont {Chipps}}, \bibinfo {author}
  {\bibfnamefont {J.}~\bibnamefont {Cizewski}}, \bibinfo {author}
  {\bibfnamefont {N.}~\bibnamefont {Curtis}}, \bibinfo {author} {\bibfnamefont
  {K.~L.}\ \bibnamefont {Jones}}, \bibinfo {author} {\bibfnamefont
  {T.}~\bibnamefont {Munoz-Britton}}, \bibinfo {author} {\bibfnamefont {S.~D.}\
  \bibnamefont {Pain}}, \bibinfo {author} {\bibfnamefont {N.}~\bibnamefont
  {Soi\ifmmode~\acute{c}\else \'{c}\fi{}}}, \bibinfo {author} {\bibfnamefont
  {C.}~\bibnamefont {Wheldon}}, \bibinfo {author} {\bibfnamefont {G.~L.}\
  \bibnamefont {Wilson}}, \ and\ \bibinfo {author} {\bibfnamefont {V.~A.}\
  \bibnamefont {Ziman}},\ }\href {\doibase 10.1103/PhysRevC.90.054324}
  {\bibfield  {journal} {\bibinfo  {journal} {Phys. Rev. C}\ }\textbf {\bibinfo
  {volume} {90}},\ \bibinfo {pages} {054324} (\bibinfo {year}
  {2014})}\BibitemShut {NoStop}%
\end{thebibliography}%

\end{document}